\documentclass[varenna]{cimento}

\usepackage[english]{babel}
\usepackage{latexsym}
\usepackage{graphics}
\usepackage{epsfig}
\usepackage{color}
\usepackage{bm}
\usepackage{amsmath}
\usepackage{amssymb}

\usepackage{accents}

\usepackage{mathtools}

\usepackage{braket}

\usepackage[normalem]{ulem}
\usepackage{color}

\newcommand{\1}{\uparrow}
\newcommand{\2}{\downarrow}

\title{Beyond-mean-field effects in mixtures: few-body and many-body aspects}
\author{D.~S.~Petrov}
\institute{LPTMS, CNRS, Univ. Paris Sud, Universit\'e Paris-Saclay, 91405 Orsay, France}

\begin{document}
\maketitle
\begin{abstract}
The discovery of ultracold dilute liquids has significantly elevated our interest in various phenomena which go under the name of beyond-mean-field (BMF) physics. In these lecture notes we give an elementary introduction to the quantum stabilization and liquefaction of a collapsing weakly interacting Bose-Bose mixture. A detailed derivation of the leading BMF correction, also known as the Lee-Huang-Yang (LHY) term, in this system is presented in a manner suitable for further generalizations and extensions. Although the LHY term is a nonanalytic function of the density $n$, under certain conditions the leading BMF correction becomes analytic and can be expanded in integer powers of $n$, effectively introducing three-body and higher-order interactions. We discuss why and how well the Bogoliubov approach can predict these few-body observables.

\end{abstract}

\section{Liquid versus dilute liquid}

There is nothing unusual in liquids. We deal with them every day. Liquid state and gas-liquid transition are well described by the van der Waals theory. The self-bound property is due to an attractive finite-range force acting between atoms or molecules of the liquid. To extract a particle from the liquid one has to work against this force at the interface. This extraction energy is called the particle emission threshold. On the other hand, the interparticle interaction is characterized by a repulsive core, which does not allow the system to collapse. The bulk density of the liquid (called saturation density) results from a compromise between the attractive and repulsive parts of the potential. Usual liquids are thus dense in the sense that the mean interparticle separation is comparable to the range of the interaction. In mathematical form these arguments can be represented by writing the energy per particle for densities close to the saturation density $n_0$ as  
\begin{equation}\label{EnergyPerParticleGeneral}
\frac{E}{N}\approx U(n_0)+|\kappa|(n-n_0)^2.
\end{equation} 

Adjective {\it quantum} is usually added to the term liquid to emphasize that we are dealing with liquids in the regime of quantum degeneracy. In this context we should mention extensive studies of liquid He droplets~\cite{Barranco} and large nuclei~\cite{Vautherin,Bender}. Although quantum degenerate, these liquids are dense. Since they lack any small expansion parameter they either require an {\it ab initio} analysis or phenomenological density-functional theories, optimized for describing particular static or dynamic properties of experimental interest. 

The term {\it dilute quantum droplet} appeared in the work of Bulgac~\cite{Bulgac} who pointed out that Eq.~(\ref{EnergyPerParticleGeneral}) could be realized in a dilute Bose gas with two-body attraction and three-body repulsion. In this case the energy density reads
\begin{equation}\label{TwoThreeBodyEnergy}
\frac{E}{V}=g_2n^2/2+g_3n^3/6
\end{equation}       
and the saturation density $n_0=-(3/2)g_2/g_3$ can be made arbitrarily low for small $g_2$ and large $g_3$. The proposal of Ref.~\cite{Bulgac} is based on a resonant Efimov enhancement of $g_3$ implying significant inelastic losses. Engineering an elastic $g_3$ while keeping the two-body interaction weak (remember that we need small $g_2$) is a nontrivial task. In these lecture notes we will explain that this effective three-body interaction can come from the leading BMF correction in certain weakly interacting bosonic models (weak interaction is the key requirement for suppressing inelastic losses in free-space bosonic gases).

The low-density expansion of the energy of a dilute gas is not necessarily in integer powers of $n$. Indeed, the celebrated LHY formula for the energy of a weakly interacting three-dimensional single-component Bose condensate reads~\cite{LHY}
\begin{equation}\label{LHYscalar}
E/V=(g_2n^2/2)(1+128\sqrt{na^3}/15\sqrt{\pi}+...),
\end{equation}
where $n$ is the density and $a>0$ and $g_2=4\pi\hbar^2a/m$ are, respectively, the scattering length and coupling constant characterizing the interparticle interaction. Equation~(\ref{LHYscalar}) contains the mean-field (MF) energy and the LHY correction. These are the two leading-order terms in the expansion of the energy density in the small gas parameter $\sqrt{na^3}$. The LHY term originates from zero-point energies of the Bogoliubov excitations and is thus intrinsically quantum. It is also universal in the sense that it depends only on the two-body scattering length (in three dimensions the shape of the interaction potential becomes important at the next order). Quite naturally the experimental observation of this fundamental BMF effect comes from the field of ultra-cold gases \cite{Altmeyer2007,Shin2008,Navon2010,Papp2008,Pollack2009,Navon2011}, where the parameter $\sqrt{na^3}$ and, therefore, the relative contribution of the LHY term, can be enhanced by using Feshbach resonances \cite{ChinRMP}. However, Eq.~(\ref{LHYscalar}) does not predict any liquid phase. Indeed, for repulsive $g_2$ there is no minimum in $E/N$ since both MF and LHY terms are repulsive. For $a<0$, Eq.~(\ref{LHYscalar}) is not applicable, but even if we take it at its face value, the LHY term becomes imaginary and does not lead to any physically sensible results.

In Ref.~\cite{PetrovLHY} we noted that in a Bose-Bose mixture one can independently control the MF and LHY terms and make them comparable to each other without ever leaving the weakly-interacting regime. In particular, the liquid phase can be achieved when the MF term, proportional to $n^2$, is negative and the LHY term, proportional to $n^{5/2}$, is positive. Because of its steeper density scaling, the quantum LHY repulsion neutralizes the MF attraction and stabilizes the system against collapse (quantum stabilization). The mixture can then exist as a droplet in equilibrium with vacuum without any external trapping. This finding naturally suggested a proof-of-principle experiment for observing the LHY quantum correction and such droplets were indeed observed in homonuclear~\cite{Cabrera2018,Semeghini2018,Cheiney2018} and heteronuclear~\cite{DErrico2019,Guo2021} Bose-Bose mixtures.

\section{Quantum stabilization mechanism\label{Sec:Stabilization}}

To understand the origin of the LHY correction and the phenomenon of quantum stabilization let us look at the following illustrative one-body examples. First, consider a particle of unit mass moving in the potential $U(x)$ which has a (for simplicity, global) minimum at $x_0$. The classical ground state of the Hamiltonian $p^2/2+U(x)$ obviously corresponds to a particle localized at $x=x_0$ and $p=0$ with energy $U(x_0)$. Translated to the Bose-Einstein-condensate (BEC) language, this state corresponds to the classical-field ground state, i.e., the solution of the Gross-Pitaevskii equation. 

To take into account quantum effects one can approximate $U(x)\approx U(x_0)+U''(x_0)(x-x_0)^2/2$ and use the well-known solution of the quantum-mechanical harmonic oscillator problem with the Hamiltonian 
\begin{equation}\label{HamOsc}
U(x_0)+\hat{p}^2/2+\omega^2(\hat{x}-x_0)^2/2,
\end{equation} 
where $\omega=\sqrt{U''(x_0)}$. We know that the ground state energy is now increased by the zero-point energy $\omega/2$ (here and after $\hbar=1$) with respect to the bottom of the potential well. The quantum correction to the ground state energy can thus be deduced from the classical spectrum of small oscillations around the classical ground state.

The way we have just obtained the BMF correction $\omega/2$ is a shortcut since we know the harmonic oscillator physics by hart. A more formal derivation is to diagonalize the harmonic Hamiltonian by introducing the lowering and raising operators $\hat{a}=\sqrt{\omega/2}\hat{x}+i\hat{p}/\sqrt{2\omega}$ and $\hat{a}^\dagger=\sqrt{\omega/2}\hat{x}-i\hat{p}/\sqrt{2\omega}$ with the result
\begin{equation}\label{HamOscDiag}
\hat{H}_{{\rm ho}}=\hat{p}^2/2+\omega^2\hat{x}^2/2=\omega(\hat{a}^\dagger\hat{a}+\hat{a}\hat{a}^\dagger)/2=\omega\hat{a}^\dagger\hat{a}+\omega/2.
\end{equation}

In the BEC language this ``shortcut'' corresponds to the following. We expand the time-dependent classical-field Gross-Pitaevskii equation around the classical ground state up to linear terms. The resulting equations (they are at least two: for real and imaginary parts of the fluctuations) are called the Bogoliubov-de Gennes equations. They give eigenfunctions and eigenfrequencies of the Bogoliubov modes. The LHY correction is essentially the sum of the zero-point energies of these Bogoliubov modes (see, however, Sec.~\ref{SecQuantAn}). It corresponds to the second term on the right-hand side of Eq.~(\ref{LHYscalar}). The quantity $\omega^2$ in the one-body example roughly corresponds to the interaction strength $\propto g_2$ in the BEC case.

The collapse instability in the scalar BEC, which occurs when $g_2$ becomes negative, can thus be compared to changing the sign of $U''(x_0)$ (or $\omega^2$) in the one-body example. In this case the minimum at $x_0$ turns into maximum and the classical solution becomes unstable. However, we understand that ``switching on'' quantum mechanics does not stabilize this classically unstable solution.

\begin{figure}[hbtp]
\begin{center}
\includegraphics[clip,width=0.6\columnwidth]{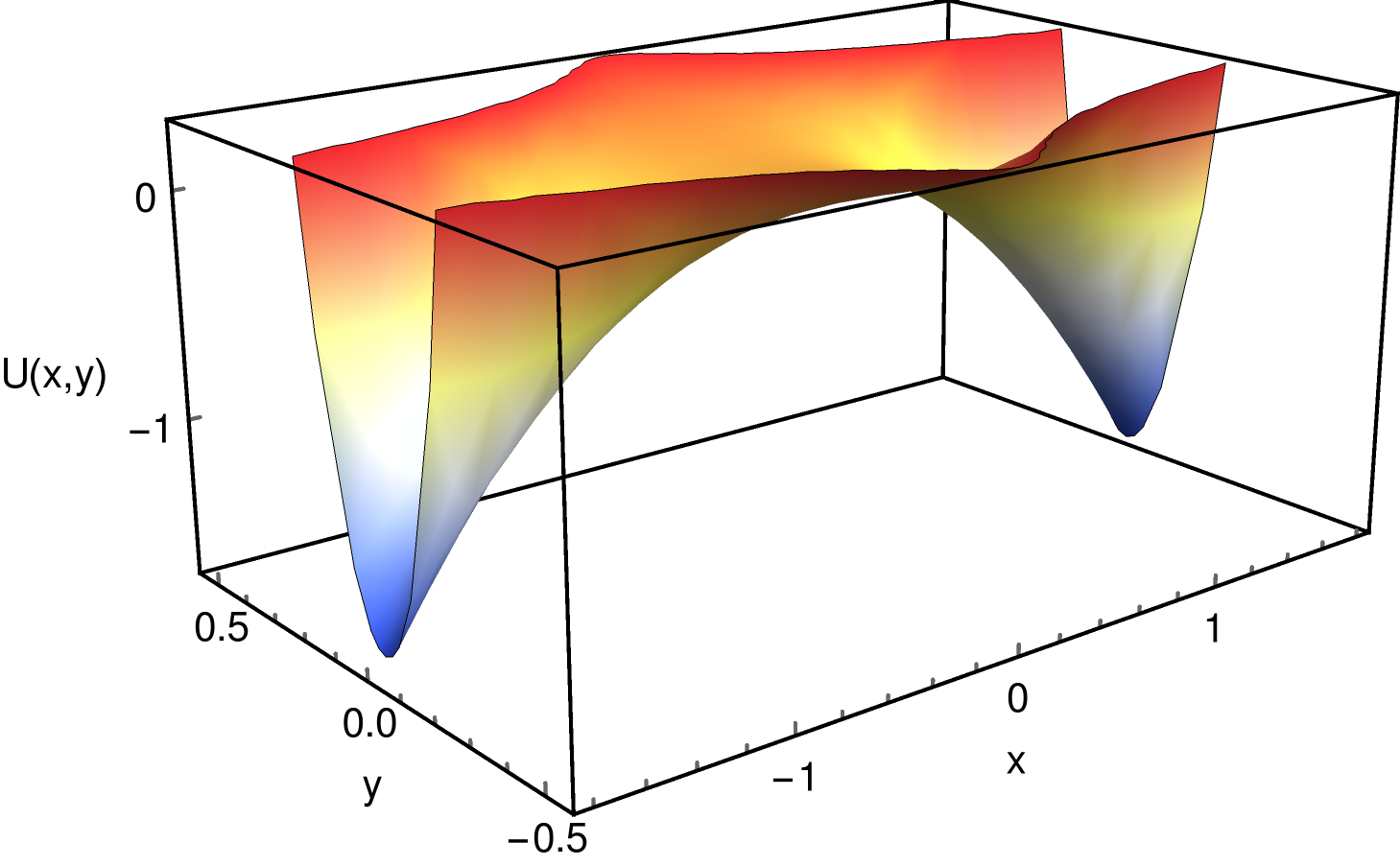}
\end{center}
\caption{
The potential given by Eq.~(\ref{Uxy}) has a form of a valley in the $x$ direction with the walls becoming steeper as $|x|$ increases. The quantum-mechanical zero-point energy of oscillations along $y$ provides an additional effective confining potential along $x$ such that the particle is trapped. By contrast, this potential has no minimum and, therefore, no stable classical solution.}
\label{FigUxy}
\end{figure}

Can a classically unstable system be stable from the quantum-mechanical viewpoint? The answer is positive. Consider, for instance, a two-dimensional particle of unit mass in the potential
\begin{equation}\label{Uxy}
U(x,y)=-x^2/2+y^2(1+6x^2+2x^4+2x^2y^2)
\end{equation}
shown in Fig.~\ref{FigUxy}. The potential Eq.~(\ref{Uxy}) has no minimum at finite $x$ and $y$. A classical particle would thus roll down the $y=0$ valley to $x=\pm \infty$. By contrast, one can check that the quantum-mechanical problem is solved by the wave function $\Psi(x,y)\propto \exp[-(1+x^2)(1/2+y^2)]$~\footnote{To come up with this example we first guessed the wave function and then differentiated it to get the potential. In this manner one can ``engineer'' many examples of this kind.}. What happens is that the confinement along $y$ becomes steeper with increasing $|x|$ such that the quantum-mechanical zero-point energy of oscillations along $y$ provides an effective trapping potential along $x$. 

This example demonstrates how a ``classical'' instability in one degree of freedom ($x$) can get stabilized by quantum-mechanical effects in the other degree of freedom ($y$). Here, the system is chosen such that the solution is analytic. A more general case is more difficult to analyse. An important obstacle is related to the fact that we usually like to start with a well-defined classical ground state and then add quantum corrections. What to do in practice, if there is no classical ground state and the exact solution is not available? One strategy can be guessed from the above example, if the motion along $x$ is much slower than the oscillations along $y$. In this case, for any fixed $x$ one finds the classical ground state energy and quantum correction to the motion in the $y$ direction. This quantity is then used as an effective $x$-dependent potential $U_{eff}(x)$ when we solve for the ``slow'' motion along $x$ (one says that the motion along $y$ is integrated out). The behavior of the bare model, limited to the classical solution in the $y$ direction, and the effective model, which takes into account quantum fluctuations, can be qualitatively different. This is what happens in a Bose-Bose mixture where the density channel is unstable with respect to ``slow'' collapse, but ``fast'' quantum-mechanical fluctuations in the spin channel stabilize the system. 

\section{BMF effects in a Bose-Bose mixture}\label{SecBoseBoseMixture}

The Bogoliubov theory is a powerful approach for describing quantum effects in a many-body system in the vicinity of its classical ground state. In particular, for bosonic systems, the leading BMF correction (in these lectures we call it the LHY term even though the system is not a scalar Bose gas) is obtained by diagonalizing the quadratic Bogoliubov Hamiltonian and is (related to) the half sum of the frequencies of its normal modes. In this section we present all ingredients needed for solving the problem of a self-bound weakly interacting Bose-Bose mixture. We hope that the level of detail will make these notes also useful to tackle other cases (more spin components, Rabi-coupled or spin-orbit coupled mixtures, etc.) In fact, we will briefly discuss the Rabi-coupled case in Sec.~\ref{SecRC}. As a historical remark, the LHY correction for the Bose-Bose mixture was calculated by Larsen in 1963~\cite{Larsen}. Interestingly, this work was motivated by the phenomenon of immiscibility of $^4$He and $^3$He, which is a Bose-Fermi mixture.

Consider a Bose-Bose mixture governed by the Hamiltonian
\begin{equation}\label{Ham}
\hat{H}=\sum_{\sigma,\bf k}\left(\frac{k^2}{2m_\sigma}-\mu_\sigma\right)\hat{a}_{\sigma,{\bf k}}^\dagger \hat{a}_{\sigma,{\bf k}}+\frac{1}{2}\sum_{\sigma,\sigma',{\bf k}_1,{\bf k}_2,{\bf q}} \hat{a}_{\sigma,{\bf k}_1+{\bf q}}^\dagger \hat{a}_{\sigma',{\bf k}_2-{\bf q}}^\dagger U_{\sigma\sigma'}({\bf q}) \hat{a}_{\sigma,{\bf k}_1}\hat{a}_{\sigma',{\bf k}_2},
\end{equation}
where $U_{\sigma\sigma'}$ are interaction potentials and $\sigma=\1,\2$. The wave vector ${\bf k}$ labels single-particle states in the usual manner. A convenient normalization of these states is provided by employing the concept of a ``large unit'' volume where $\sum_{\sigma,\bf k}=\sum_{\sigma}\int d^Dk/(2\pi)^D$~\cite{LL3}. In fact, Eq.~(\ref{Ham}) describes the Hamiltonian density in the grand-canonical description, where the chemical potentials $\mu_\sigma$ are Lagrange multipliers, needed to satisfy constraints on the average densities of the components. The expectation value of $\hat{H}$ is thus the grand potential density and we will use $E$ to denote the energy density. The creation and annihilation operators $\hat{a}_{\sigma,{\bf k}}^\dagger$ and  $\hat{a}_{\sigma,{\bf k}}$ are the raising and lowering operators of harmonic oscillators which live at the points $\{\sigma,{\bf k}\}$ of the configurational space. In the second-quantized many-body case these operators are not related to coordinates and momenta of real bosons. Nevertheless, Eq.~(\ref{Ham}) is just a multi-dimensional version of a single-particle Hamiltonian, and we can apply the same procedure of calculating the LHY correction as in Sec.~\ref{Sec:Stabilization}.

\subsection{Classical analysis}

Let us first see how far we can go along the ``shortcut'' route. To this end we pass to the classical version of Eq.~(\ref{Ham}) by  replacing all operators by $c$-numbers. We have explicitly
\begin{equation}\label{ClHam}
H=\sum_{\sigma,\bf k}\left(\frac{k^2}{2m_\sigma}-\mu_\sigma\right)a_{\sigma,{\bf k}}^* a_{\sigma,{\bf k}}+\frac{1}{2}\sum_{\sigma,\sigma',{\bf k}_1,{\bf k}_2,{\bf q}} a_{\sigma,{\bf k}_1+{\bf q}}^* a_{\sigma',{\bf k}_2-{\bf q}}^* U_{\sigma\sigma'}({\bf q}) a_{\sigma,{\bf k}_1}a_{\sigma',{\bf k}_2}.
\end{equation}
The corresponding classical equation of motion, also called the time-dependent Gross-Pitaevskii (GP) equation, reads
\begin{equation}\label{ClEqMot}
i\dot{a}_{\sigma,{\bf k}}=\frac{\delta H}{\delta a^*_{\sigma,{\bf k}}}=\left(\frac{k^2}{2m_\sigma}-\mu_\sigma\right)a_{\sigma,{\bf k}}+\sum_{\sigma',{\bf k}_1,{\bf q}} U_{\sigma\sigma'}({\bf q})a_{\sigma',{\bf k}_1}^* a_{\sigma',{\bf k}_1+{\bf q}} a_{\sigma,{\bf k}-{\bf q}}.
\end{equation}
Equation~(\ref{ClEqMot}), being complex, is equivalent to two real equations (for $x$ and $p$ in the single-particle example). The second equation (for the conjugate field $a^*_{\sigma,{\bf k}}$) is just the Hermitian conjugate of Eq.~(\ref{ClEqMot}).

The ground state is stationary, i.e., it satisfies Eq.~(\ref{ClEqMot}) with vanishing left-hand side. If the interaction $U$ is weak, a good starting point to search for the ground state is the condensate at ${\bf k}=0$, i.e., we take finite $a_{\sigma,0}$ and $a_{\sigma,0}^*$ and vanishing $a_{\sigma,{\bf k}}=a_{\sigma,{\bf k}}^*=0$ for ${\bf k}\neq 0$. One can check that this configuration solves Eq.~(\ref{ClEqMot}), if the condensate mode satisfies
\begin{equation}\label{HomGP}
\sum_{\sigma'} U_{\sigma\sigma'}(0)|a_{\sigma',0}|^2 = \mu_\sigma.
\end{equation}
The phases of $a_{\sigma,0}$ can be arbitrary [manifestation of the U(1) symmetry]. This gives rise to two Goldstone modes (since we deal with two components). To move on we should thus pick up some values for these two phases, following the symmetry-breaking BEC theory. This approach is very practical although it implies that the particle number is uncertain and can be fixed only on average. Differences between the symmetry-breaking and particle-conserving approaches have been discussed by Castin~\cite{CastinLesHouches}. From the viewpoint of calculating the LHY correction these subtleties are not very important. Without loss of generality we assume that $a_{\sigma,0}$ are real and we introduce condensate densities $n_\sigma=a_{\sigma,0}^2$. 

\begin{figure}[hbtp]
\begin{center}
\includegraphics[clip,width=1\columnwidth]{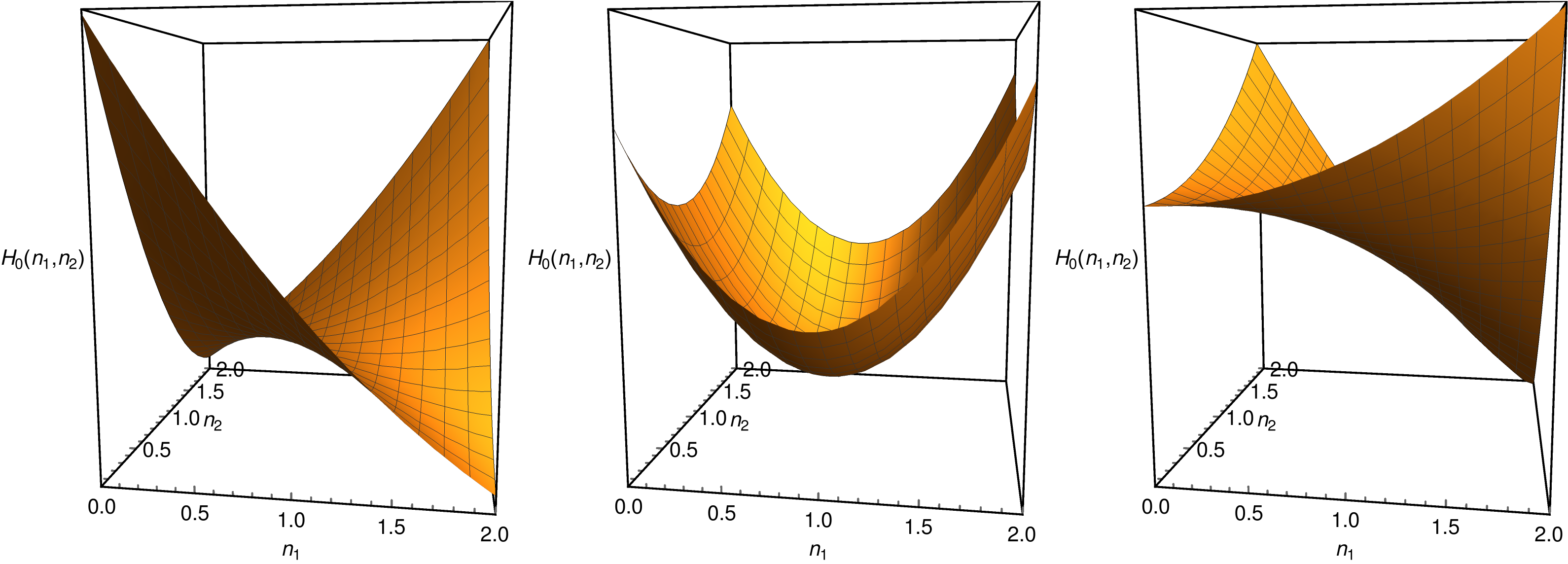}
\end{center}
\caption{
The MF grand-canonical Hamiltonian Eq.~(\ref{H0}) versus $n_1$ and $n_2$ for a Bose-Bose mixture in the regime of phase separation (left panel), stability (middle panel), and collapse (right panel).}
\label{FigGrandPotentials}
\end{figure}

Equation~(\ref{HomGP}) results from extremalizing (not minimizing) Eq.~(\ref{ClHam}) and is thus not necessarily the ground state. We exclude more exotic ground states like, for instance, supersolid. Even without breaking the translational symmetry the homogeneous solution may be mechanically unstable with respect to collapse or phase separation (low-momentum instability). This can be seen by going back to the Hamiltonian (\ref{ClHam}) with neglected finite-momentum amplitudes
\begin{equation}\label{H0}
H_0=-\sum_\sigma \mu_\sigma n_{\sigma}+\frac{1}{2}\sum_{\sigma\sigma'}U_{\sigma\sigma'}(0)n_{\sigma} n_{\sigma'}.
\end{equation}
The state given by Eq.~(\ref{HomGP}) can be a minimum, a maximum, or a saddle point of Eq.~(\ref{H0}) depending on whether the matrix $U_{\sigma\sigma'}(0)$ [Hessian of $H_0(n_{\1},n_{\2})$] is characterized, respectively, by two positive eigenvalues, by two negative eigenvalues, or by eigenvalues of different signs. One can check that the system is stable when the following three conditions are satisfied. $U_{\1\1}(0)>0$, $U_{\2\2}(0)>0$, and $U_{\1\2}^2(0)<U_{\1\1}(0)U_{\2\2}(0)$. Assuming that the first two conditions are true (otherwise, the components collapse individually) we face two possible mechanical instabilities: when $U_{\1\2}(0)>\sqrt{U_{\1\1}(0)U_{\2\2}(0)}$ the species phase separate as they repel each other too strongly. By contrast, when $U_{\1\2}(0)<-\sqrt{U_{\1\1}(0)U_{\2\2}(0)}$ they attract each other so strongly that they collapse, the repulsive intraspecies interactions being unable to stabilize the system. For illustration, in Fig.~\ref{FigGrandPotentials} we show the MF grand canonical Hamiltonian $H_0$ as a function of $n_{\1}$ and $n_{\2}$ in the regimes of phase separation (left panel), MF stability (middle), and collapse (right).

More generally, that the chosen classical state is a minimum (at least local) can be understood by studying small-amplitude excitations. In the stable case all frequencies should be real (remember, in the single-particle example the ground state is stable for $\omega^2>0$). A saddle point would mean that there are excitations with complex frequencies. In this case one says that the system is dynamical unstable.

Let us now study small-amplitude excitations around the homogeneous condensate solution satisfying Eq.~(\ref{HomGP}). To this end we introduce $\delta a_{\sigma,{\bf k}}=a_{\sigma,{\bf k}}-\sqrt{n_\sigma}\delta_{{\bf k},0}$ and linearize Eq.~(\ref{ClEqMot}) for small $\delta a_{\sigma,{\bf k}}$ obtaining
\begin{equation}\label{BdG1}
i\delta \dot{a}_{\sigma,{\bf k}}=\frac{k^2}{2m_\sigma} \delta a_{\sigma,{\bf k}}+\sum_{\sigma'} U_{\sigma\sigma'}({\bf k})\sqrt{n_\sigma n_{\sigma'}}(\delta a_{\sigma',{\bf k}}+ \delta a_{\sigma',-{\bf k}}^*).
\end{equation}
These equations (for different $\sigma$ and ${\bf k}$) can be closed by writing similar equations for $\delta a_{\sigma,-{\bf k}}^*$. Explicitly,
\begin{equation}\label{BdG2}
-i\delta \dot{a}_{\sigma,-{\bf k}}^*=\frac{k^2}{2m_\sigma} \delta a_{\sigma,-{\bf k}}^*+\sum_{\sigma'} U_{\sigma\sigma'}({\bf k})\sqrt{n_\sigma n_{\sigma'}}(\delta a_{\sigma',{\bf k}}+ \delta a_{\sigma',-{\bf k}}^*).
\end{equation}
Equations~(\ref{BdG1}) and (\ref{BdG2}) are called the Bogoliubov-de Gennes (BdG) equations. For finite $k$ we can use the ansatz
\begin{eqnarray}\label{TimeDepAnsatz}
\delta a_{\sigma,{\bf k}}(t)&=&\delta a_{\sigma,{\bf k}}e^{-iE_{\bf k} t},\nonumber\\
\delta a_{\sigma,-{\bf k}}^*(t)&=&\delta a_{\sigma,-{\bf k}}^*e^{-iE_{\bf k} t}.
\end{eqnarray}
The problem then reduces to a linear matrix equation. For the two-component case we have  
\begin{equation}\label{BdGFinal}
{\cal{L}}\begin{pmatrix}
a_{\1,{\bf k}} \\
a_{\2,{\bf k}} \\
a_{\1,-{\bf k}}^* \\
a_{\2,-{\bf k}}^*
\end{pmatrix}=E_{\bf k}\begin{pmatrix}
a_{\1,{\bf k}} \\
a_{\2,{\bf k}} \\
a_{\1,-{\bf k}}^* \\
a_{\2,-{\bf k}}^*
\end{pmatrix},
\end{equation}
where the Bogoliubov matrix reads
\begin{equation}\label{BogMatrix}
\cal{L}=\begin{pmatrix}
\frac{k^2}{2m_\1}+U_{\1\1}({\bf k})n_\1 & U_{\1\2}({\bf k})\sqrt{n_\1 n_\2} & U_{\1\1}({\bf k})n_\1 & U_{\1\2}({\bf k})\sqrt{n_\1 n_\2} \\

U_{\1\2}({\bf k})\sqrt{n_\1 n_\2} & \frac{k^2}{2m_\2}+U_{\2\2}({\bf k})n_\2 & U_{\1\2}({\bf k})\sqrt{n_\1 n_\2} & U_{\2\2}({\bf k})n_\2 \\

-U_{\1\1}({\bf k})n_\1 & -U_{\1\2}({\bf k})\sqrt{n_\1 n_\2} & -\frac{k^2}{2m_\1}-U_{\1\1}({\bf k})n_\1 & -U_{\1\2}({\bf k})\sqrt{n_\1 n_\2} \\

-U_{\1\2}({\bf k})\sqrt{n_\1 n_\2} & -U_{\2\2}({\bf k})n_\2 & -U_{\1\2}({\bf k})\sqrt{n_\1 n_\2} & -\frac{k^2}{2m_\2}-U_{\2\2}({\bf k})n_\2
\end{pmatrix}.
\end{equation}
The spectrum of (\ref{BogMatrix}) is composed of four eigenvalues $E_{+,{\bf k}}$, $E_{-,{\bf k}}$, $-E_{+,{\bf k}}$, and $-E_{-,{\bf k}}$, where
\begin{equation}\label{Epm}
E_{\pm,{\bf k}}=\sqrt{\frac{\omega_\1^2({\bf k})+\omega_\2^2({\bf k})}{2}\pm\sqrt{\frac{[\omega_\1^2({\bf k})-\omega_\2^2({\bf k})]^2}{4}+\frac{U^2_{\1\2}({\bf k})n_\1n_\2k^4}{m_\1m_\2}}},
\end{equation}
and $\omega_\sigma({\bf k})=\sqrt{U_{\sigma\sigma}({\bf k})n_\sigma k^2/m_\sigma+(k^2/2m_\sigma)^2}$ are the Bogoliubov spectra for the individual components.

That we have symmetric positive and negative eigenvalues is a mathematical artifact related to the special form of the matrix Eq.~(\ref{BogMatrix}) (see Appendix). One can easily check that the eigenstates corresponding to $E_{s,{\bf k}}$ and $-E_{s,-{\bf k}}$ ($s=\pm$) are mathematically and physically identical. Concerning the energy of the mode $\left\{s,{\bf k}\right\}$, one can calculate it by substituting the corresponding solution of Eq.~(\ref{BdGFinal}) into the classical Hamiltonian Eq.~(\ref{ClHam}) and subtracting the ground state energy Eq.~(\ref{H0}). For small oscillation amplitudes $\propto a_{\sigma,{\bf k}}$ the resulting excitation energy is quadratic in the amplitudes and is also proportional to $E_{s,{\bf k}}^2$. This can be compared to the energy $\omega^2x_{\rm max}^2/2$ of the classical oscillator Eq.~(\ref{HamOscDiag}). Therefore, on the classical level only the square of $E_{\pm,{\bf k}}$ is important, not the sign. 

Excitations with ${\bf k}=0$ have to be treated separately as in this case the ansatz Eq.~(\ref{TimeDepAnsatz}) is not suitable. The solution obtained directly from Eqs.~(\ref{BdG1}) and (\ref{BdG2}) reads ${\rm Re} \delta a_{\sigma,0}(t)={\rm const}$ and ${\rm Im}\delta a_{\sigma,0}(t)\propto t$. This solution eventually leaves the small-amplitude regime and we have to get back to Eq.~(\ref{ClEqMot}). Physically, finite values of ${\rm Re} \delta a_{\sigma,0}$ mean that we are dealing with ``wrong'' condensate densities, in the sense that they are not optimal for the given chemical potentials. The phases of these condensates thus change slightly faster or slower than $-\mu_\sigma t$. This is why we see linear divergence of ${\rm Im}\delta a_{\sigma,0}(t)$. The best way to deal with these modes is just to switch to the new chemical potentials. It may also be useful to mention that the zero-momentum modes are described by the Hamiltonian of the type $\hat{a}^\dagger \hat{a}+\hat{a} \hat{a}^\dagger + \hat{a}^\dagger \hat{a}^\dagger +\hat{a}\hat{a}$. By rotating $\hat{a}=(\hat{x}+i\hat{p})/\sqrt{2}$ the system reduces to an oscillator with infinite mass $\hat{H}(\hat{x},\hat{p})=2\hat{x}^2$, evolution of which is similar (and rather trivial) in quantum and classical cases. In what follows we pay no more attention to the mode ${\bf k}=0$ assuming $\delta a_{\sigma,0}=0$ and, later, $\delta \hat{a}_{\sigma,0}=0$.  

Going back to finite ${\bf k}$ and assuming that $U_{\sigma\sigma'}({\bf k})\rightarrow U_{\sigma\sigma'}(0)$ when ${\bf k}\rightarrow 0$, the Bogoliubov modes (\ref{Epm}) at small ${\bf k}$ are phononic $E_{\pm}({\bf k})\approx c_{\pm}k$ with sound velocities satisfying
\begin{eqnarray}\label{c2}
c_\pm^2&=&\frac{U_{\1\1}(0)n_\1/m_\1+U_{\2\2}(0)n_\2/m_\2}{2}\nonumber\\
&\pm&\sqrt{\left[\frac{U_{\1\1}(0)n_\1/m_\1+U_{\2\2}(0)n_\2/m_\2}{2}\right]^2+[U_{\1\2}^2(0)-U_{\1\1}(0)U_{\2\2}(0)]\frac{n_\1 n_\2}{m_\1 m_\2}}.
\end{eqnarray}
One can see that in the stable regime discussed earlier, i.e., for $U_{\1\2}^2(0)<U_{\1\1}(0)U_{\2\2}(0)$, both sound velocities are real (and positive). Otherwise, $c_{-}$ is imaginary meaning that the lower $E_{-,{\bf k}}$ branch is unstable for sufficiently small ${\bf k}$.

This low-momentum behavior of the excitations is consistent with the stability analysis of the MF energy density Eq.~(\ref{H0}) independent of the mass ratio, although the overall (finite-momentum) behavior of the Bogoliubov branches can be quite different in the mass-balanced and mass-imbalanced cases. In the mass-balanced case, the two branches can be cast in the form of the usual Bogoliubov dispersions $E_{\pm}({\bf k})=\sqrt{c_\pm({\bf k})^2 k^2+k^4/4m^2}$. Here, we have introduced momentum dependent quantities $c_\pm({\bf k})$ obtained by replacing $U_{\sigma\sigma'}(0)\rightarrow U_{\sigma\sigma'}({\bf k})$ in Eq.~(\ref{c2}). From the viewpoint of excitations, a mass-balanced mixture, even with arbitrary $U_{\sigma\sigma'}({\bf k})$, is thus equivalent to two decoupled scalar Bose gases. This decoupling works only in the mass-balanced case. For $m_\1 \neq m_\2$ the two branches can interwine in a rather peculiar manner (for instance, they can exhibit an avoided crossing at finite $k$). 

Let us move on and discuss the LHY correction. Following the ``shortcut'' intuition, the LHY term should equal half the sum of the excitation frequencies. We face a few difficulties here. First, we do not know which sign to choose for the frequencies, $+E_{\sigma,{\bf k}}$ or $-E_{\sigma,{\bf k}}$. The formal procedure is discussed in Appendix. It is basically equivalent to choosing the branch adiabatically connected to the correct excitation frequency in the noninteracting limit. In the concrete example of the mixture we take $+E_{\sigma,{\bf k}}$ as this branch is connected to $+k^2/2m_\sigma$, which is the correct excitation energy in the noninteracting limit.

The second difficulty is that the integral $\sum_{\pm}\int E_{\pm,{\bf k}}d^Dk/(2\pi)^D$ diverges at large momentum. This is particularly annoying in the noninteracting case, where instead of a vanishing LHY correction we have a divergent integral. However, the problem is so apparent that one can easily see where it comes from. Indeed, our ``shortcut'' procedure for calculating the zero-point energy is exact for the oscillator Hamiltonian in the symmetric form $(\hat{a}^\dagger \hat{a}+\hat{a} \hat{a}^\dagger)/2$, but fails for the normally-ordered $\hat{a}^\dagger \hat{a}$, where it still predicts 1/2 instead of 0 for the ground state energy. Since both of these quantum Hamiltonians have the same classical limit $a^*a$ the problem does not come from  classical mechanics. Nevertheless, the ``shortcut'' approach is still useful. We just have to keep track of the ordering of the operators in the original Hamiltonian.

\subsection{Quantum analysis}\label{SecQuantAn}

Let us go back to the quantum Hamiltonian Eq.~(\ref{Ham}) and, in contrast to what we did before, we now keep ``quantum'' hats over the finite-momentum field operators assuming that only the condensate mode is classical and real $\hat{a}_{\sigma,0}=\sqrt{n_\sigma}$ satisfying Eq.~(\ref{HomGP}). We then expand the Hamiltonian in powers of $\hat{a}_{\sigma,{\bf k}\neq 0}$. We have explicitly
\begin{equation}\label{Expansion}
\hat{H}=H_0+\hat{H}_2+\hat{H}_3+\hat{H}_4,
\end{equation} 
where $H_0$ is given by Eq.~(\ref{H0}),
\begin{equation}\label{H2}
\begin{aligned}
\hat{H}_2&=\sideset{}{'}\sum_{\sigma,\bf k}\left[\frac{k^2}{2m_\sigma}-\mu_\sigma+\sum_{\sigma'}U_{\sigma\sigma'}(0)n_{\sigma'}\right] \hat{a}_{\sigma,{\bf k}}^\dagger \hat{a}_{\sigma,{\bf k}}\\
&+\frac{1}{2}\sideset{}{'}\sum_{\sigma,\sigma',{\bf k}}U_{\sigma\sigma'}({\bf k})\sqrt{n_{\sigma} n_{\sigma'}}(\hat{a}_{\sigma,{\bm k}}^\dagger\hat{a}_{\sigma',-{\bm k}}^\dagger+\hat{a}_{\sigma,{\bm k}}\hat{a}_{\sigma',-{\bm k}}+2\hat{a}_{\sigma,{\bm k}}^\dagger\hat{a}_{\sigma',{\bm k}}),
\end{aligned}
\end{equation}
\begin{equation}\label{H3}
\hat{H}_3=\sideset{}{'}\sum_{\sigma,\sigma',{\bm k}_1,{\bm k}_2}U_{\sigma{\sigma'}}({\bf k}_1)\sqrt{n_{\sigma}}\hat{a}_{\sigma',{\bm k}_1+{\bm k}_2}^\dagger(\hat{a}_{\sigma,{\bm k}_1}+\hat{a}_{\sigma,-{\bm k}_1}^\dagger)\hat{a}_{\sigma',{\bm k}_2}
\end{equation}
and
\begin{equation}\label{H4}
\hat{H}_4=\frac{1}{2}\sideset{}{'}\sum_{\sigma,\sigma',{\bf k}_1,{\bf k}_2,{\bf q}} \hat{a}_{\sigma,{\bf k}_1+{\bf q}}^\dagger \hat{a}_{\sigma',{\bf k}_2-{\bf q}}^\dagger U_{\sigma\sigma'}({\bf q}) \hat{a}_{\sigma,{\bf k}_1}\hat{a}_{\sigma',{\bf k}_2}.
\end{equation}  
Primes in Eqs.~(\ref{H2}-\ref{H4}) indicate that the corresponding sum excludes terms involving creation or annihilation operators of condensate particles. 

The Bogoliubov approximation consists of neglecting $\hat{H}_3$ and $\hat{H}_4$. This is legal for sufficiently small fluctuations, which is the case if interactions are weak (remember that the pure condensate is the ground state of an ideal gas). Note that the remaining part, $H_0+\hat{H}_2$, quadratic in the field operators, if we decided to follow the ``shortcut'' approach, would give exactly the same BdG equations (\ref{BdG1}) and (\ref{BdG2}) and would also lead to the same ``issue'' associated with the diverging integral over the Bogoliubov energies. Keeping the quantum nature of the fields the quadratic Hamiltonian can be diagonalized as follows. By using the commutation rules $\hat{a}_{\sigma,{\bm k}}^\dagger\hat{a}_{\sigma',{\bm k}}=(1/2)(\hat{a}_{\sigma,{\bm k}}^\dagger\hat{a}_{\sigma',{\bm k}}+\hat{a}_{\sigma,{\bm k}}\hat{a}_{\sigma',{\bm k}}^\dagger)-\delta_{\sigma\sigma'}/2$ we rewrite Eq.~(\ref{H2}) in the symmetric form,
\begin{equation}\label{BogHamSymm}
\hat{H}_2=-\frac{1}{2}\sideset{}{'}\sum_{\sigma,\bf k}\left[\frac{k^2}{2m_\sigma}+U_{\sigma\sigma}({\bf k})n_{\sigma}\right]+\frac{1}{2}\sideset{}{'}\sum_{\bf k}(\hat{a}_{\1,{\bf k}}^\dagger \hat{a}_{\2,{\bf k}}^\dagger \hat{a}_{\1,-{\bf k}} \hat{a}_{\2,-{\bf k}})
\bar{\cal{L}}
\begin{pmatrix}
\hat{a}_{\1,{\bf k}} \\
\hat{a}_{\2,{\bf k}} \\
\hat{a}_{\1,-{\bf k}}^\dagger \\
\hat{a}_{\2,-{\bf k}}^\dagger
\end{pmatrix},
\end{equation}
where
\begin{equation}\label{Lbar}
\bar{\cal{L}}=
\begin{pmatrix}
\frac{k^2}{2m_\1}+U_{\1\1}({\bf k})n_\1 & U_{\1\2}({\bf k})\sqrt{n_\1 n_\2} & U_{\1\1}({\bf k})n_\1 & U_{\1\2}({\bf k})\sqrt{n_\1 n_\2} \\

U_{\1\2}({\bf k})\sqrt{n_\1 n_\2} & \frac{k^2}{2m_\2}+U_{\2\2}({\bf k})n_\2 & U_{\1\2}({\bf k})\sqrt{n_\1 n_\2} & U_{\2\2}({\bf k})n_\2 \\

U_{\1\1}({\bf k})n_\1 & U_{\1\2}({\bf k})\sqrt{n_\1 n_\2} & \frac{k^2}{2m_\1}+U_{\1\1}({\bf k})n_\1 & U_{\1\2}({\bf k})\sqrt{n_\1 n_\2} \\

U_{\1\2}({\bf k})\sqrt{n_\1 n_\2} & U_{\2\2}({\bf k})n_\2 & U_{\1\2}({\bf k})\sqrt{n_\1 n_\2} & \frac{k^2}{2m_\2}+U_{\2\2}({\bf k})n_\2
\end{pmatrix}.
\end{equation}
We now understand that in spite of the fact that the Bogoliubov Hamiltonian (\ref{BogHamSymm}) features $\bar{\cal{L}}$, its diagonalization should somehow be based on the eigenstates and eigenvalues of $\cal{L}$ given by Eq.~(\ref{BogMatrix}). Indeed, one can show (see more details in Appendix) that Eq.~(\ref{BogHamSymm}) is diagonalized by the Bogoliubov transformation
\begin{equation}\label{BogTrans}
\begin{pmatrix}
\hat{a}_{\1,{\bf k}} \\
\hat{a}_{\2,{\bf k}} \\
\hat{a}_{\1,-{\bf k}}^\dagger \\
\hat{a}_{\2,-{\bf k}}^\dagger
\end{pmatrix}
=\hat{S}
\begin{pmatrix}
\hat{b}_{+,{\bf k}} \\
\hat{b}_{-,{\bf k}} \\
\hat{b}_{+,-{\bf k}}^\dagger \\
\hat{b}_{-,-{\bf k}}^\dagger
\end{pmatrix},
\end{equation}
where the columns of the matrix $\hat{S}$ are (properly normalized) eigenvectors of (\ref{BogMatrix}) corresponding, respectively, to the eigenvalues $E_{+,{\bf k}}$, $E_{-,{\bf k}}$, $-E_{+,{\bf k}}$, and $-E_{-,{\bf k}}$. The quadratic form in Eq.~(\ref{BogHamSymm}) then transforms according to
\begin{equation}\label{}
\hat{S}^\dagger
\bar{\cal{L}}
\hat{S}=
\begin{pmatrix}
E_{+,{\bf k}} &&&0\\
&E_{-,{\bf k}}&&\\
&&E_{+,{\bf k}}&\\
0&&&E_{-,{\bf k}}
\end{pmatrix}
\end{equation}
and the whole quadratic part becomes (we use $\hat{b}_{s,{\bf k}}^\dagger\hat{b}_{s,{\bf k}}+\hat{b}_{s,{\bf k}}\hat{b}_{s,{\bf k}}^\dagger=1+2\hat{b}_{s,{\bf k}}^\dagger\hat{b}_{s,{\bf k}}$)
\begin{equation}\label{BogHamSymmDiag}
\hat{H}_2=\frac{1}{2}\sideset{}{'}\sum_{\bf k}\left\{\sum_{s=\pm}E_{s,{\bf k}}-\sum_{\sigma=\1,\2}\left[\frac{k^2}{2m_{\sigma}}+U_{\sigma\sigma}({\bf k})n_{\sigma}\right]\right\}+\sum_{s=\pm}\sum_{\bf k}E_{s,{\bf k}}\hat{b}_{s,{\bf k}}^\dagger\hat{b}_{s,{\bf k}}.
\end{equation}
We can finally write down the ground-state energy in the Bogoliubov approximation 
\begin{equation}\label{EMFLHY}
E_{\rm MF+LHY}=\frac{1}{2}\sum_{\sigma\sigma'}U_{\sigma\sigma'}(0)n_{\sigma} n_{\sigma'}+\frac{1}{2}\sideset{}{'}\sum_{\bf k}\left\{\sum_{s=\pm}E_{s,{\bf k}}-\sum_{\sigma=\1,\2}\left[\frac{k^2}{2m_{\sigma}}+U_{\sigma\sigma}({\bf k})n_{\sigma}\right]\right\}.
\end{equation}
The sum over momenta is now converging thanks to the last term in Eq.~(\ref{EMFLHY}). We remind that this term comes from the correct account of the order of operators in the original quantum Hamiltonian. This is the only element for which we need to know the structure of the quantum Hamiltonian. Generalization of this procedure to generic Hamiltonians is more or less obvious. The key ingredients for calculating the LHY term are the convenient reordering of the operators in $\hat{H}_2$, as we do in Eq.~(\ref{BogHamSymm}), and finding the spectrum of small-amplitude excitations around the classical ground state. In principle, we do not even need to know the wavefunctions of these excitations.

With Eq.~(\ref{EMFLHY}) we are one step away from analyzing the influence of the LHY term on the mechanical stability of the system. In the next section we will discuss how to treat the momentum summation in Eq.~(\ref{EMFLHY}) in the dilute limit and we will obtain closed-form expressions for the ground-state energy. However, before we proceed, let us mention one point important for the stability analysis. The derivation leading to Eqs.~(\ref{BogHamSymmDiag}) and (\ref{EMFLHY}) is valid only in the case of a stable MF state, i.e., when the Bogoliubov energies are real. Equation~(\ref{BogHamSymmDiag}) is then a representation of the quadratic Bogoliubov Hamiltonian in terms of decoupled oscillators of the type $\hat{p}^2+\hat{x}^2$. When $U_{\1\2}^2(0)$ exceeds $U_{\1\1}(0)U_{\2\2}(0)$ the branch $E_-({\bf k})$ becomes imaginary (at small $k$). These excitations can still be represented as independent oscillators, this time of the type $\hat{p}^2-\hat{x}^2$, which can no longer be reduced to $\hat{b}^\dagger\hat{b}$ (see Appendix). However, as we have mentioned in the end of Sec.~\ref{Sec:Stabilization}, the problem can still be treated when the typical rate of instability dynamics is much slower than the frequencies of stable modes dominating the LHY term. For the Bose-Bose mixture this happens when the difference between $U_{\1\2}^2(0)$ and $U_{\1\1}(0)U_{\2\2}(0)$ is much smaller than each of these terms. In this case we can adiabatically integrate out high-frequency high-momentum modes and arrive at an effective low-frequency low-momentum theory where the effect of the higher modes is represented by the LHY term. As we will see below, the contribution of the low-frequency modes to the LHY integral is subleading. The low-energy theory thus essentially reduces to the classical GPE (the so-called modified GPE) where the LHY term, coming from high-energy modes, is taken into account in the local-density approximation. The key point is that if the MF instability is weak (or slow), the LHY term can stabilize it in spite of the fact that this term is of higher-order in the gas parameter. Here, it can also be useful to remember the analogy with a particle in the two-dimensional potential Eq.~(\ref{Uxy}). Along the valley, the classical potential $U(x,0)=-x^2/2$ is unstable, but if we integrate out the motion in the $y$ direction and take into account the corresponding zero-point energy, which scales like $\omega(x)/2\approx x^2$ at large $x$, we see that the renormalized theory becomes stable.

\subsection{Renormalization, effective potentials, low-energy scattering observables}
\label{Subsec:Renorm}

In the dilute regime the interaction between atoms $\sigma$ and $\sigma'$ is characterized by a single parameter, the scattering length $a_{\sigma\sigma'}$, or, equivalently, by the coupling constant $g_{\sigma\sigma'}=2\pi a_{\sigma\sigma'}/m_{\sigma\sigma'}$, where we introduce the reduces masses $m_{\sigma\sigma'}=m_\sigma m_{\sigma'}/(m_\sigma+m_{\sigma'})$. The coupling constant is the ground state interaction shift for a pair of atoms in a unit volume. To the leading order we have $g_{\sigma\sigma'}=U_{\sigma\sigma'}(0)$. Can we just replace $U_{\sigma\sigma'}({\bf k})$ by $g_{\sigma\sigma'}$ in the Bogoliubov theory? To answer this question let us expand the term in the curly brackets in Eq.~(\ref{EMFLHY}) at large $k$. To the leading order in $1/k$ we have
\begin{equation}\label{IntAsymptote}
\sum_{s=\pm}E_{s,{\bf k}}-\sum_{\sigma=\1,\2}\left[\frac{k^2}{2m_{\sigma}}+U_{\sigma\sigma}({\bf k})n_{\sigma}\right]\approx 
-\sum_{\sigma\sigma'}\frac{2m_{\sigma\sigma'} U_{\sigma\sigma'}^2({\bf k})n_\sigma n_\sigma'}{k^2}.
\end{equation}
If we replace $U_{\sigma\sigma'}({\bf k})\rightarrow g_{\sigma\sigma'}$, in two and three dimensions this would lead to an ultraviolet divergence in Eq.~(\ref{EMFLHY}) [remember that $\sum_{\bf k}=\int d^Dk/(2\pi)^D$]. 

The large-$k$ behavior of $U_{\sigma\sigma'}(k)$ is thus important. We also see from Eq.~(\ref{IntAsymptote}) that the large-momentum part of this integral gives a contribution proportional to $n^2$ to the energy density, i.e., it renormalizes the two-body interaction. In three dimensions one formally rewrites Eq.~(\ref{EMFLHY}) as the sum of 
\begin{equation}\label{EMF}
E_{\rm MF}=\frac{1}{2}\sum_{\sigma\sigma'}n_{\sigma} n_{\sigma'}\left[U_{\sigma\sigma'}(0)-\sideset{}{'}\sum_{\bf k}\frac{2m_{\sigma\sigma'}U_{\sigma\sigma'}^2({\bf k})}{k^2}\right],
\end{equation}
and 
\begin{equation}\label{LHY}
E_{\rm LHY}=\frac{1}{2}\sideset{}{'}\sum_{\bf k}\left\{\sum_{s=\pm}E_{s,{\bf k}}-\sum_{\sigma=\1,\2}\left[\frac{k^2}{2m_{\sigma}}+U_{\sigma\sigma}({\bf k})n_{\sigma}\right]+\sum_{\sigma\sigma'}\frac{2m_{\sigma\sigma'} U_{\sigma\sigma'}^2({\bf k})n_\sigma n_\sigma'}{k^2}\right\},
\end{equation}
where the term (\ref{EMF}) is conventionally called the MF energy and (\ref{LHY}) the LHY correction, in spite of the fact that a significant part of the second-order contribution $\propto U^2$ has been transferred from one to the other. This is consistent with the fact that the expansions in terms of the gas parameter $\sqrt{na^3}$ and in terms of powers of the bare interaction $U$ are different. Let us agree that the MF energy is the exact two-body interaction energy $g_{\sigma\sigma'}$ summed over all pairs. Then, in Eq.~(\ref{EMF}) one recognizes the first two terms in the perturbative expansion 
\begin{equation}\label{gfromU}
g_{\sigma\sigma'}\approx U_{\sigma\sigma'}(0)-\sideset{}{'}\sum_{\bf k}\frac{2m_{\sigma\sigma'}U_{\sigma\sigma'}^2({\bf k})}{k^2}.
\end{equation}
Note that the integral is converging since we do not replace $U_{\sigma\sigma'}({\bf k})$ by a constant. However, we need to ensure that the perturbative expansion is legal. The usual restriction on the Born series expansion at low energy reads $|U_{\sigma\sigma'}|\ll 1/(\mu_{\sigma\sigma'}\kappa)$, where $\kappa$ is the typical range of the potential in momentum space~\cite{LL3}. For realistic van der Waals potentials this condition is not satisfied since they are too deep. In such cases the realistic potential is replaced by an effective potential characterized by the same collisional properties at low energies, but which can be treated perturbatively. For not too exotic potentials the only parameter that we need is $a_{\sigma\sigma'}$ or $g_{\sigma\sigma'}$. As far as the LHY term (\ref{LHY}) is concerned, the integral converges, dominated by momenta on the order of the healing momentum $\sim \sqrt{mgn}$ (for estimates, we introduce $m\sim m_\1\sim m_\2$, $n\sim n_\1\sim n_\2$, and $g\sim |g_{\sigma\sigma'}|$), which we assume to be much smaller than $\kappa$. We can thus make the replacement $U_{\sigma\sigma'}(k)\rightarrow U_{\sigma\sigma'}(0)\approx g_{\sigma\sigma'}$ in Eq.~(\ref{LHY}) and in the Bogoliubov spectra (\ref{Epm}). 

Let us now briefly discuss low-dimensional cases. In one dimension the integral in Eq.~(\ref{EMFLHY}), with $U_{\sigma\sigma'}(k)$ replaced by a constant, converges at large $k$. The replacement $U_{\sigma\sigma'}(k)\rightarrow U_{\sigma\sigma'}(0)=g_{\sigma\sigma'}$ is thus possible if the typical momentum dominating the integral (which is the healing momentum defined above) is much smaller than the momentum range of the potential $\kappa$. In other words, we approximate the actual finite-range potential by a delta potential $g_{\sigma\sigma'}\delta(x)$, the scattering on which is well defined in one dimension. The link between the many-body energy expansion Eq.~(\ref{EMFLHY}) and low-energy scattering properties is then provided by the relation  between the coupling constant and the scattering length $g_{\sigma\sigma'}=-1/(m_{\sigma\sigma'}a_{\sigma\sigma'})$.

In two dimensions replacing $U_{\sigma\sigma'}({\bf k})$ by a constant would lead to a logarithmic divergence in Eq.~(\ref{EMFLHY}) and renormalizing $U_{\sigma\sigma'}({\bf q})$, like in three dimensions, could make sense. However, the second-order integrals in Eqs.~(\ref{EMF}) and (\ref{gfromU}) then diverge at low momentum. The link between the many-body Bogoliubov expansion and two-body scattering observables can be constructed by introducing the momentum cutoff and passing to effective potentials $U_{\sigma\sigma'}({\bf k})=g_{\sigma\sigma'}={\rm const}\ll 1$ for $|{\bf k}|<\kappa$ and $U_{\sigma\sigma'}({\bf k})=0$ for $|{\bf k}|>\kappa$. The coupling constants $g_{\sigma\sigma'}$ and the cutoff $\kappa$  are related to the two-dimensional scattering lengths $a_{\sigma\sigma'}>0$ by $g_{\sigma\sigma'}=2\pi/[m_{\sigma\sigma'}\ln(2m_{\sigma\sigma'}\epsilon_{\sigma\sigma'}/\kappa^2)]$, where $\epsilon_{\sigma\sigma'}=2e^{-2\gamma}/(m_{\sigma\sigma'}a_{\sigma\sigma'}^2)$ and $\gamma$ is Euler's constant. This relation ensures that at low collision energy $E\ll \kappa^2/m$ the scattering $t$-matrix (analog of $g$ in other dimensions) behaves as $t_{\sigma\sigma'}(E)\approx 2\pi/[m_{\sigma\sigma'}\ln(-\epsilon_{\sigma\sigma'}/E)]$ \cite{Popov1971,Jackiw1991} consistent with the Born series expansion $t_{\sigma\sigma'}(E)\approx g_{\sigma\sigma'}[1-m_{\sigma\sigma'}g_{\sigma\sigma'}\ln(-\kappa^2/2m_{\sigma\sigma'}E)/2\pi+...]$. One can see that the perturbation series in terms of $m|t|\ll 1$ and $m|g|\ll 1$ are equivalent as long as $\kappa^2/m$ is larger but not exponentially larger than the typical interaction energy $E$, which is the product of the density $n$ and the $t$-matrix (with the logarithmic accuracy one can simply use $E\sim n/m$). An appropriate value of $\kappa$ can always be found in the weakly-interacting regime where the scattering lengths are exponentially small (repulsion) or large (attraction) compared to the mean interparticle separation.

In this approach the cutoff is formally necessary, although it should not influence the final result. This is to say that Eq.~(\ref{EMFLHY}), which is an expansion of the energy to the second order in $m|g|\ll 1$, should not be sensitive to $\kappa$ to this order. Indeed, let us fix $a_{\sigma\sigma'}$ and compare $E_{\rm MF+LHY}$ calculated for $\kappa$ and $\tilde{\kappa}$. The difference between the corresponding coupling constants, directly related to the difference of the MF parts of Eq.~(\ref{EMFLHY}), can be written as
\begin{equation}\label{gtildeminusg}
\tilde{g}_{\sigma\sigma'}-g_{\sigma\sigma'}=\frac{2\pi}{m_{\sigma\sigma'}\ln(2m_{\sigma\sigma'}\epsilon_{\sigma\sigma'}/\tilde{\kappa}^2)}-\frac{2\pi}{m_{\sigma\sigma'}\ln(2m_{\sigma\sigma'}\epsilon_{\sigma\sigma'}/\kappa^2)}\approx \frac{m_{\sigma\sigma'}g_{\sigma\sigma'}^2}{2\pi}\ln\frac{\kappa^2}{\tilde{\kappa}^2},  
\end{equation}
which is a second-order quantity and, therefore, with the same accuracy, we can replace $g$ by $\tilde{g}$ in the last term. We see that the MF part varies in the second order as we pass from $\kappa$ to $\tilde{\kappa}$. One can also check [by integrating Eq.~(\ref{IntAsymptote}) over momenta from $\kappa$ to $\tilde{\kappa}$] that the corresponding change in the LHY part of Eq.~(\ref{EMFLHY}) compensates the variation of the MF part. The final result is thus indeed insensitive to variations of $\kappa$ (to the second order). A simple manner of removing the cutoff dependence from the final equations is to choose $\kappa$ such that the expression for the LHY correction formally vanishes. The energy then formally contains only the MF part. However, the coupling ``constant'' $g$ is now density dependent since $\kappa$ is on the order of the inverse healing length $\sqrt{gn}$ [see, for example, Eq.~(\ref{EMFLHYRes})].

\subsection{Mass-balanced mixtures in the Bogoliubov approximation}

For short-range effective potentials just discussed, the momentum integration in Eq.~(\ref{EMFLHY}) can be performed analytically and $E_{\rm MF+LHY}$ can be written explicitly in terms of elliptic functions (see Supplemental Material of Ref.~\cite{NaidonPetrov}) simplifying the analysis of thermodynamic properties of the mixture. We will now discuss the mass-balanced case where the density dependence of $E_{\rm MF+LHY}$ is particularly simple.

For equal masses we explicitly have (we set $m_\1=m_\2=1$) 
\begin{equation}\label{EMFLHYShortRange}
E_{\rm MF+LHY}=\frac{1}{2}\sum_{\sigma\sigma'}g_{\sigma\sigma'}n_\sigma n_{\sigma'}+\frac{1}{2}\sum_{s=\pm} \left\{
\begin{matrix}
\vspace{4mm}
\sum_{{\bf k}}[E_{s,k}-k^2/2-c_s^2+c_{s}^4/k^2], & D=3,\\
\vspace{4mm}
\sum_{|{\bf k}|<\kappa}[E_{s,k}-k^2/2-c_s^2], & D=2,\\
\sum_{k}[E_{s,k}-k^2/2-c_s^2], & D=1,
\end{matrix}
\right.
\end{equation}
where $E_{\pm,k}=\sqrt{c_\pm^2 k^2+k^4/4}$ and Eq.~(\ref{c2}) now reduces to
\begin{equation}\label{cpm}
c_{\pm}^2=\frac{g_{\1\1}n_\1+g_{\2\2}n_\2\pm\sqrt{(g_{\1\1}n_\1-g_{\2\2}n_\2)^2+4g_{\1\2}^2 n_\1 n_\2}}{2}.
\end{equation}
The momentum integration in Eq.~(\ref{EMFLHYShortRange}) gives
\begin{equation}\label{EMFLHYRes}
E_{\rm MF+LHY}=\frac{1}{2}\sum_{\sigma\sigma'}g_{\sigma\sigma'}n_\sigma n_{\sigma'}+\sum_{s=\pm} \left\{
\begin{matrix}
\vspace{4mm}
\frac{8}{15\pi^2} c_{s}^5, & D=3,\\
\vspace{4mm}
\frac{1}{8\pi}c_{s}^4  \ln(c_s^2\sqrt{e}/\kappa^2), & D=2,\\
-\frac{2}{3\pi}c_{s}^3, & D=1.
\end{matrix}
\right.
\end{equation}

In contrast to the purely MF energy density, which is a quadratic form of $n_\1$ and $n_\1$, the LHY correction makes Eq.~(\ref{EMFLHYRes}) a more peculiar function of the densities. The Hessian of the energy density now varies as a function of $n_1$ and $n_2$, which means that the system can, in principle, be stable or unstable at different points of the $\{n_\1,n_\2\}$ space. In other words, $E_{\rm MF+LHY}(n_\1,n_\2)$ can have convex (stable) and concave (unstable) regions. It turns out that this behavior can lead to liquefaction, when we are close to the collapse threshold, or to partial miscibility, when we are close to the regime of phase separation. 

\begin{figure}[hbtp]
\begin{center}
\includegraphics[clip,width=1\columnwidth]{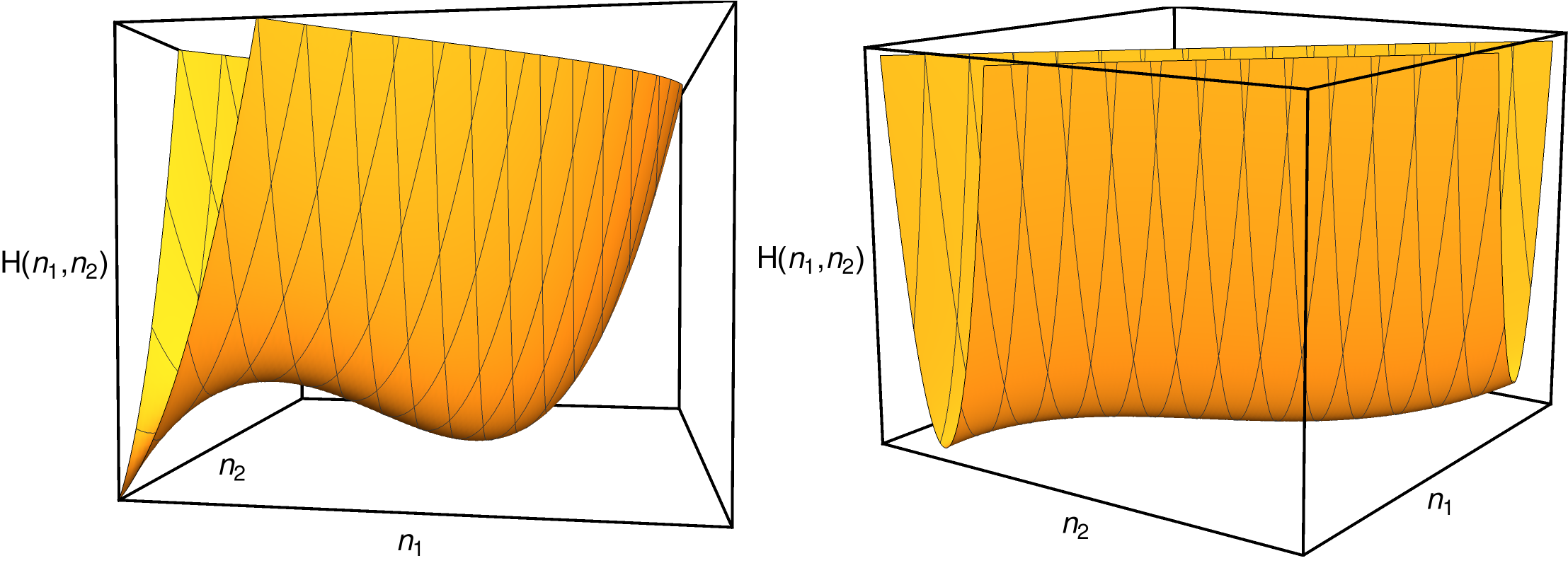}
\end{center}
\caption{
The grand potential Eq.~(\ref{GrandPotentialMFLHY}) including the MF and LHY contributions versus $n_1$ and $n_2$ for a Bose-Bose mixture in the regimes of liquid (left panel) and partial phase separation (right panel).}
\label{FigLiquidAndBubble}
\end{figure}

From the thermodynamical viewpoint, these phenomena are associated with first-order phase transitions and with coexistence of two phases. The self-trapped liquid is (by definition) in equilibrium with vacuum. There is a transition curve in the $\{\mu_\1,\mu_\2\}$ space where the grand potential 
\begin{equation}\label{GrandPotentialMFLHY}
H(n_\1,n_\2)=E_{\rm MF+LHY}(n_\1,n_\2)-\mu_\1 n_\1-\mu_\2 n_\2
\end{equation} 
has two degenerate minima as a function of $n_\1$ and $n_\2$: vacuum with $n_\1=n_\2=0$ and matter with finite $\{n_\1,n_\2\}$. In fact, these two points correspond to vanishing pressure (which is minus the grand potential density). The left panel of Fig.~\ref{FigLiquidAndBubble} shows an example corresponding to the matter-vacuum phase transition when the plane $\mu_\1 n_\1+\mu_\2 n_\2$ touches the three-dimensional surface $E_{\rm MF+LHY}(n_\1,n_\2)$ at two points. 

Partial miscibility is another BMF phenomenon, which is due to the fact that the LHY correction is not quadratic in densities. Indeed, the MF quadratic form $E_{\rm MF}(n_\1,n_\2)$ is either purely convex or purely concave implying respectively purely miscible or purely immiscible regimes close to the phase separation threshold. By contrast, the LHY correction can render $E_{\rm MF+LHY}(n_\1,n_\2)$ partially convex and partially concave, meaning that the components can mix, but not at any proportion. In the right panel of Fig.~\ref{FigLiquidAndBubble} we show an example of $H(n_\1,n_\2)$, which corresponds to the first-order phase transition between a mixed phase and a pure phase.

These phenomena rely on a competition between the MF and LHY terms in Eq.~(\ref{EMFLHYRes}). These are the first two leading terms in powers of the weak-interaction parameter $\eta\ll 1$, which scales in different dimensions as $\eta_{D=3}\propto \sqrt{g^3n}$, $\eta_{D=2}\propto g$, and $\eta_{D=1}\propto \sqrt{g/n}$ (again, we take $g_{\1\1}\sim g_{\2\2}\sim g_{\1\2}\sim g$ and $n_\1 \sim n_\2 \sim n$).
Since these parameters should be small for the validity of the Bogoliubov approximation how is it then possible that the higher-order LHY term can compete with the MF term? The answer to this question is that close to the collapse or to the phase separation threshold the MF term does become small in the sense that one of the eigenvalues of the matrix $g_{\sigma\sigma'}$ becomes small. The corresponding eigenvector designates a direction in the $n_{\1}n_\2$-plane, along which the system is ``soft'' and sensitive to the LHY correction, whereas in the perpendicular direction the system's behavior is still governed by the dominant MF term. This separation of scales makes the analysis of the phases, which consists of minimizing the grand potential Eq.~(\ref{GrandPotentialMFLHY}) with respect to the densities, a two-step process. The phenomenon of partial miscibility for interaction-imbalanced and mass-imbalanced mixtures close to the phase-separation threshold has been analyzed theoretically in Ref.~\cite{NaidonPetrov}. Here, we focus on the collapse threshold and discuss the liquid phase. 

\subsection{Dilute liquid in mass-balanced mixtures}

Let us introduce $\delta g = g_{\1\2}+\sqrt{g_{\1\1}g_{\2\2}}$ and consider the case $|\delta g|/g\sim \eta\ll 1$. In this regime the ``soft'' part of the MF term is comparable to the LHY correction since $\delta g n^2\sim E_{\rm LHY}\sim \eta gn^2$. Diagonalizing the MF quadratic form at $\delta g=0$ suggests rotating the $n_\1 n_\2$-plane as follows.
\begin{eqnarray}
n_+&=&\frac{\sqrt{g_{\1\1}}n_\1-\sqrt{g_{\2\2}}n_\2}{\sqrt{g_{\1\1}+g_{\2\2}}},\label{nplus}\\
n_-&=&\frac{\sqrt{g_{\2\2}}n_\1+\sqrt{g_{\1\1}}n_\2}{\sqrt{g_{\1\1}+g_{\2\2}}}.\label{nminus}
\end{eqnarray}
We also introduce the corresponding chemical potentials
\begin{eqnarray}
\mu_+&=&\frac{\sqrt{g_{\1\1}}\mu_\1-\sqrt{g_{\2\2}}\mu_\2}{\sqrt{g_{\1\1}+g_{\2\2}}},\label{muplus}\\
\mu_-&=&\frac{\sqrt{g_{\2\2}}\mu_\1+\sqrt{g_{\1\1}}\mu_\2}{\sqrt{g_{\1\1}+g_{\2\2}}}.\label{muminus}
\end{eqnarray}
The grand potential density (for finite $\delta g$) in the new variables reads
\begin{eqnarray}
H(n_+,n_-)&=&\frac{g_{\1\1}+g_{\2\2}}{2}n_+^2-\mu_+ n_+\nonumber\\
 &\hspace{-10mm}+&\hspace{-5mm}\delta g\frac{(\sqrt{g_{\2\2}}n_-+\sqrt{g_{\1\1}}n_+)(\sqrt{g_{\1\1}}n_--\sqrt{g_{\2\2}}n_+)}{g_{\1\1}+g_{\2\2}}+E_{\rm LHY}-\mu_- n_-,\label{GProtated}
\end{eqnarray}
where we place the leading-order terms ($\propto gn^2$) in the first line and the next-order ones ($\propto gn^2\eta$) in the second line. Not to exceed the accuracy of the Bogoliubov approximation (we cannot go higher than $\eta g n^2$) we should set $\delta g=0$ in $E_{\rm LHY}$ (recall that $\delta g\sim \eta g$), which amounts to replacing $c_-$ by $0$ and $c_+^2$ by 
\begin{equation}\label{cp0}
c_+^2|_{\delta g=0}=g_{\1\1}n_\1+g_{\2\2}n_\2=\sqrt{g_{\1\1}g_{\2\2}}\frac{\sqrt{g_{\1\1}}+\sqrt{g_{\2\2}}}{\sqrt{g_{\1\1}+g_{\2\2}}}n_-+\frac{g_{\1\1}^{3/2}-g_{\2\2}^{3/2}}{\sqrt{g_{\1\1}+g_{\2\2}}}n_+.
\end{equation} 
in Eq.~(\ref{EMFLHYRes}).

According to the hierarchy of powers of $\eta$, the first step is to minimize Eq.~(\ref{GProtated}) on the MF level, i.e., we minimize its first line with respect to $n_+$. This gives
\begin{equation}\label{dMF}
n_+=\frac{\mu_+}{g_{\1\1}+g_{\2\2}}.
\end{equation} 
We can now minimize on the LHY level (terms $\propto \eta gn^2$). Although $n_+$ does get corrected by $\delta n_+\sim \eta n_{+}$, this leads to a correction $\sim gn^2\eta^2$ in Eq.~(\ref{GProtated}), which can be neglected on the LHY level. In the second step we thus arrive at the problem of minimizing $H$ with respect to $n_-$ at fixed $n_+$ given by Eq.~(\ref{dMF}).

The two-step procedure just described is analogous to the Born-Oppenheimer-type separation of variables into fast and slow. In the case of a mixture close to the collapse threshold fast degrees of freedom are fluctuations of $n_+$ [for $g_{\1\1}=g_{\2\2}$ this would be spin, see Eq.~(\ref{nplus})] and slow correspond to fluctuations of $n_-$ (total density). 

Let us now discuss the self-bound phase in equilibrium with vacuum. In this case the chemical potentials $\mu_\1$ and $\mu_\2$ should be negative. One can also speak about self-bound bubbles for positive chemical potentials (or one of the chemical potentials), but these are objects immersed in a gas, i.e., not in vacuum (see, for instance, Ref.~\cite{Tengstrand2022}). The ``vacuum'' requirement leads to an important restriction on $\mu_+$ and on the population imbalance in the liquid. To see this we note that, in order to be the same as in vacuum, $H$ should be zero on both MF and LHY levels. Therefore, $\mu_-$ should be on the order of $\delta g n$. From Eq.~(\ref{muminus}) one then understands that $\mu_\1$ and $\mu_\2$ individually are also $\sim \delta g n$. Otherwise, one of them would have to become positive. We thus arrive at the conclusion that $\mu_+ \sim \delta g n$. Then, according to Eq.~(\ref{dMF}) $n_+\sim (\delta g/g) n \ll n$. This means that the ratio $n_\1/n_\2$ is ``locked'' to the value $\sqrt{g_{\2\2}/g_{\1\1}}$. Deviations of order $\delta g/g$ of this ratio are possible. Otherwise, one of the chemical potentials crosses zero and the liquid can no longer bind additional particles of that component. 

The ``vacuum'' restriction simplifies Eqs.~(\ref{GProtated}) and (\ref{cp0}), where we can now set $n_+=0$. The analysis of the liquid thus reduces to the minimization of 
the function
\begin{equation}
H(n_+=0,n_-)=\delta g_-n_-^2-\mu_- n_-+\left\{
\begin{matrix}
\vspace{4mm}
\frac{8}{15\pi^2} (g_-n_-)^{5/2}, & D=3,\\
\vspace{4mm}
\frac{1}{8\pi}(g_-n_-)^2  \ln(g_-n_-\sqrt{e}/\kappa^2), & D=2,\\
-\frac{2}{3\pi}(g_-n_-)^{3/2}, & D=1,
\end{matrix}
\right.\label{GProtatedFinal}
\end{equation}
where for brevity we introduce $\delta g_-=\delta g \sqrt{g_{\1\1}g_{\2\2}}/(g_{\1\1}+g_{\2\2})$ and $g_-=\sqrt{g_{\1\1}g_{\2\2}}(\sqrt{g_{\1\1}}+\sqrt{g_{\2\2}})/\sqrt{g_{\1\1}+g_{\2\2}}$.

If it were not for the symbol $\delta$ in the first term on the right-hand side, Eq.~(\ref{GProtatedFinal}) would be the grand potential, written in the Bogoliubov approximation, for a scalar Bose gas with density $n_-$ and coupling constant $g_-$. In the mixture we are able to tune the MF term (or, better say, its ``soft'' part) independently of the LHY correction, which is dominated by the ``stiff'' $+$ mode. 

To continue the analysis of the liquid phase, we should minimize $H(n_-)$ and also require that $H$ vanishes at the minimum. One can show that in the particular case of Eq.~(\ref{GProtatedFinal}) (which vanishes at $n_-=0$) these two conditions are equivalent to minimization of $H(n_-)/n_-$. The value of $n_-$ at the minimum is the saturation density. In three dimensions the minimum is found for $\delta g<0$ and the saturation density equals
\begin{equation}\label{SatDens3D}
n_-^{(3D)}=\frac{25\pi^4}{16}\frac{{\delta g}_-^2}{g_-^5}.
\end{equation}
Note that the gas parameter of the liquid $\sqrt{g^3n}\sim |\delta g|/g\ll 1$ making the inequality $|\delta g|/g\ll 1$ the only validity condition of the theory.

Liquefaction of mass-balanced Bose-Bose mixtures in low dimensions has been analyzed in Ref.~\cite{PetrovAstrakharchik2016}. In two dimensions the minimization of $H(n_-)/n_-$ leads to the equation
\begin{equation}\label{SatDens2DInterm}
\delta g_-+\frac{1}{8\pi}g_-^2\ln\frac{g_- n_- e^{3/2}}{\kappa^2}=0.
\end{equation}
The saturation density can always be found mathematically (the sign of $\delta g$ does not matter since logarithm can have any value). The liquid thus exists as long as our assumptions are satisfied. The validity conditions of the Bogoliubov theory are $|g_{\sigma\sigma'}|\ll 1$ and we also assume that $g_{\1\1}>0$, $g_{\2\2}>0$, $g_{\1\2}<0$, and $|\delta g|\ll |g_{\sigma\sigma'}|$. These conditions are equivalent to $\{\epsilon_{\1\1},\epsilon_{\2\2}\} \gg \kappa^2$ (here, $\gg$ means exponentially larger), $\epsilon_{\1\2}\ll \kappa^2$ (exponentially smaller). For simplicity we can use the freedom to chose $\kappa$ such that $\delta g = 0$. Namely, we take
\begin{equation}\label{Cutoff}
\kappa^2=\sqrt{\epsilon_{\1\2}\sqrt{\epsilon_{\1\1}\epsilon_{\2\2}}}\exp\frac{-\ln^2(\epsilon_{\1\1}/\epsilon_{\2\2})}{4\ln(\epsilon_{\1\1}\epsilon_{\2\2}/\epsilon^2_{\1\2})}. 
\end{equation}
The two-dimensional saturation density is then a cumbersome but analytic function of $\epsilon_{\sigma\sigma'}$. In the symmetric case ($\epsilon_{\1\1}=\epsilon_{\2\2}$) it reads
\begin{equation}\label{SatDens2D}
n_-^{(2D)}|_{\epsilon_{\1\1}=\epsilon_{\2\2}}=\frac{\sqrt{\epsilon_{\1\2}\epsilon_{\1\1}}}{8\sqrt{2}\pi}e^{-3/2}\ln\frac{\epsilon_{\1\1}}{\epsilon_{\1\2}}=\frac{e^{-3/2-2\gamma}}{\sqrt{2}\pi a_{\1\2}a_{\1\1}}\ln\frac{a_{\1\2}}{a_{\1\1}}.
\end{equation} 
One can check that, similarly to the three-dimensional case, all validity conditions reduce to a single one, $\ln(a_{\1\2}/a_{\1\1})\gg 1$. The only requirement for the existence of the liquid phase is thus weak intraspecies repulsions and weak interspecies attraction. The liquid then automatically chooses its density such that the usual conditions for weak interactions are satisfied, $n a_{\1\1}^2\ll 1$ and $n a_{\1\2}^2\gg 1$ (exponentially).

In the one-dimensional case the analysis of Eq.~(\ref{GProtatedFinal}) is straightforward. The saturation density is given by
\begin{equation}\label{SatDens1D}
n_-^{(1D)}=\frac{1}{9\pi^2}\frac{g_-^3}{\delta g_-^2}.
\end{equation}
One can check that the validity of the theory is ensured by a single condition $\delta g_-\ll g_-$. A peculiarity of the one-dimensional case is that the liquid exists for $\delta g_->0$ (i.e., in the regime stable from the MF viewpoint) and the saturation density diverges at small $\delta g_-$ (in one dimension the weakly interacting regime is at high densities). Interestingly, it is the attractive LHY correction which is responsible for the self trapping. For negative $\delta g_-$ both MF and LHY terms are attractive and the gas collapses (in the thermodynamic limit).

That the LHY term features a nonquadratic dependence on the densities leads to the liquefaction of the mixture in the dilute regime. Self-binding is thus a ``smoking gun'', which unequivocally attests the presence of BMF physics in a dilute ultracold gas (or, more precisely, in a dilute ultracold liquid). Quantum droplets in a mixture of two hyperfine states of $^{39}$K have been observed by Cabrera and co-workers~\cite{Cabrera2018,Cheiney2018} and by Semeghini and co-workers~\cite{Semeghini2018}. They have also been studied in heteronuclear Rb-K~\cite{DErrico2019} and Rb-Na~\cite{Guo2021} mixtures. We should also mention experimental studies of BMF effects in mixtures tuned to the regime where the MF and LHY terms are comparable, but the gas is still trapped~\cite{Skov2021}. For the vast subject of dipolar droplets we refer the reader to the lecture notes of Santos in this volume and to the recent review~\cite{PfauReview}. We just mention that for dipolar droplets the stabilization mechanism is very similar, as it also comes from a local LHY term. However, the MF physics there is more involved because of the long-range and anisotropic interaction potential.

\section{LHY correction: analytic or nonanalytic}

We now understand that the LHY term is sensitive to the structure of the spectrum of Bogoliubov quasiparticles and to their density of states. Two systems with exactly the same density dependence of the MF term can have different structures of the Bogoliubov vacuum and thus different functional forms of the LHY term. Equation~(\ref{EMFLHYRes}) is an illustration of this observation. We see that the dimension of space has a strong effect on the density dependence of the LHY term ($n^{5/2}$ for $D=3$, $n^2\ln(n)$ for $D=2$, and $n^{3/2}$ for $D=1$). Note also that this dependence is nonanalytic (exhibiting a branch-cut singularity at $n=0$), which is counterintuitive. We would expect the weak interaction expansion of the energy density to be in integer powers of $n$, when one can identify the two-body term $\propto n^2$, three-body term $\propto n^3$, etc. At least, this is what happens naturally when one looks at the problem from the few-body side (see, for instance, Refs.~\cite{Petrov2014,Grisha1D,Valiente,Grisha2D}). Integer powers of $n$ sometimes also show up in the LHY correction, in particular, in quasi-low-dimensional systems~\cite{Edler2017,Zin2021,PricoupenkoPetrov2021} or in Rabi-coupled mixtures~\cite{Cappellaro2017,Lavoine2021}. 

An interesting observation is that using the many-body Bogoliubov theory for calculating the three-body term turns out to be technically easier than solving the three-body problem directly. Indeed, we obtain the same results for the three-body coupling constant $g_3$ by using the exact few-body approach~\cite{Petrov2014} and by using the Bogoliubov theory~\cite{Lavoine2021}. The question is can we generalize this statement to other cases? Doubts originate from the fact that the three-body term turns out to be proportional to the third power of interaction, meaning that we go rather high in the perturbation theory. It is then not obvious whether the Bogoliubov theory, which neglects $\hat{H}_3$ and $\hat{H}_4$ in Eq.~(\ref{Expansion}), can properly handle these terms. Here are a few more general questions to think about. How do the expansions in terms of the Bogoliubov small parameter $\eta$, in terms of powers of the interaction strength, or in terms of the density, relate to one another? Under what conditions the LHY correction is an analytic or nonanalytic function of the density?

To answer these questions let us consider the problem of $N$ identical bosons of unit mass in a unit volume interacting by a very weak potential with Fourier transform $U({\bf k})$. We plan to compare the standard perturbation theory in powers of $U$ and the Bogoliubov approach. We start with the standard perturbation theory. Denoting the noninteracting multiparticle states by symbols with the bar $\bar{n}=\{{\bf k}_1,...,{\bf k}_N\}$,  the corresponding multiparticle energies by $\omega_{\bar{n}}$, their differences by $\omega_{\bar{n}\bar{m}}=\omega_{\bar{n}}-\omega_{\bar{m}}$, and the interaction part of the Hamiltonian by
\begin{equation}\label{Ubar}
\bar{U}=\frac{1}{2}\sum_{{\bf k}_1,{\bf k}_2,{\bf q}} \hat{a}_{{\bf k}_1+{\bf q}}^\dagger \hat{a}_{{\bf k}_2-{\bf q}}^\dagger U({\bf q}) \hat{a}_{{\bf k}_1}\hat{a}_{{\bf k}_2},
\end{equation}
energy corrections to state $\bar{n}$ read~\cite{LL3}
\begin{equation}\label{E1LL}
E_{\bar{n}}^{(1)}=\bar{U}_{\bar{n}\bar{n}},
\end{equation}  
\begin{equation}\label{E2LL}
E_{\bar{n}}^{(2)}=-{\sum}'_{\bar{m}} |\bar{U}_{\bar{m}\bar{n}}|^2/\omega_{\bar{m}\bar{n}},
\end{equation}  
\begin{equation}\label{E3LL}
E_{\bar{n}}^{(3)}={\sum}'_{\bar{k}} {\sum}'_{\bar{m}} \frac{\bar{U}_{\bar{n}\bar{m}}\bar{U}_{\bar{m}\bar{k}}\bar{U}_{\bar{k} \bar{n}}}{\omega_{\bar{m}\bar{n}}\omega_{\bar{k} \bar{n}}}-E_{\bar{n}}^{(1)}{\sum}'_{\bar{m}} \frac{|\bar{U}_{\bar{m}\bar{n}}|^2}{\omega_{\bar{m}\bar{n}}^2},
\end{equation}  
where the primes mean that the state $\bar{n}$ is excluded from the summations.

A straightforward combinatorial counting of multi-particle excited states gives the ground state energy of the $N$-body system up to third order in the form [$\binom{i}{j}$ are binomial coefficients)
\begin{equation}\label{Epert}
E^{(1)}+E^{(2)}+E^{(3)}=g_2\binom{N}{2}+g_3\binom{N}{3},
\end{equation}
where
\begin{equation}\label{g2}
g_2=U({\bf 0})-\sum_{{\bf k}} \frac{|U({\bf k})|^2}{k^2}+\sum_{{\bf q},{\bf k}} \frac{U(-{\bf q})U({\bf q}-{\bf k})U({\bf k})}{k^2q^2}
\end{equation}
and
\begin{equation}\label{g3}
g_3=6\sum_{\bf k}\frac{U^3({\bf k})}{k^4}.
\end{equation}
In Eqs.~(\ref{g2}) and (\ref{g3}), states with ${\bf k}={\bf 0}$, ${\bf q}={\bf 0}$, or ${\bf q}={\bf k}$ are excluded from the summation. Equation~(\ref{g2}) is nothing else than the first three diagrams of the ladder summation for the two-body interaction vertex. The three-body term (\ref{g3}) accounts for the following sequence of virtual excitations of three different atoms. The first and the second atoms interact with each other and get excited into states $-{\bf k}$ and ${\bf k}$, respectively. Then, the second interaction event takes place for the second and third atom. It results in the second atom getting back to the ground state and the third atom being excited to state ${\bf k}$. Finally, the first and the third atoms interact with each other both going down to the ground state. There is no four-body interaction on this level of the perturbation theory.

Now we apply the Bogoliubov theory to this problem. In the single-component case Eq.~(\ref{EMFLHY}) reads
\begin{equation}\label{EMFLHYscalar}
E_{\rm MF+LHY}=\frac{1}{2}U(0)n^2+\frac{1}{2}\sideset{}{'}\sum_{\bf k}[\sqrt{U({\bf k})nk^2+k^4/4}-k^2/2-U({\bf k})n],
\end{equation}
and let us formally expand it in powers of $U$ up to $U^3$. We obtain
\begin{equation}\label{EMFLHYscalarExp}
E_{\rm MF+LHY}=\frac{1}{2}U(0)n^2+\frac{1}{2}\sideset{}{'}\sum_{\bf k}[-n^2U^2({\bf k})/k^2+2n^3U^3({\bf k})^3/k^4].
\end{equation}
Identifying $N=n$ and assuming $N\gg 1$ we see that the Bogoliubov theory almost exactly reproduces the standard result [except for the last term in Eq.~(\ref{g2})]. We have already seen in Sec.~\ref{Subsec:Renorm} that the Bogoliubov approximation predicts the second-order renormalization of the two-body interaction. Now we see that it also predicts the three-body term in the same form as the standard perturbation theory. To understand this we note that expanding Eq.~(\ref{EMFLHYscalar}) at small $U$ is equivalent to applying the standard perturbation theory to the scalar version of the Bogoliubov Hamiltonian (\ref{H2}). This Hamiltonian allows excitations of pairs with momenta  $-{\bf k}$ and ${\bf k}$ out of the condensate, which is enough for the sequence of virtual excitations leading to $g_3$. We now also understand why the Bogoliubov theory misses the third-order renormalization of $g_2$ [the last term in Eq.~(\ref{g2})]. Indeed, this term implies that two atoms virtually excited from the condensate to states $-{\bf k}$ and ${\bf k}$ interact again and go to states $-{\bf q}$ and ${\bf q}$. Interactions between excited atoms are not allowed on the Bogoliubov level (unless they go back to the condensate). To correctly describe this term we have to take into account the quartic part $\hat{H}_4$ Eq.~(\ref{H4}). 

So, why do we not see the nonanalytic dependence of the LHY term in Eq.~(\ref{EMFLHYscalarExp})? The answer is that this expansion [as well as Eqs.~(\ref{Epert}), (\ref{g2}), and (\ref{g3})] is valid when the interaction shift per atom is smaller than the minimal excitation energy in the noninteracting gas $Un/k^2\ll 1$. In other words, the healing length $\xi\sim 1/\sqrt{Un}$ should be much larger than the system size. We can compare this with the validity condition of the general Bogoliubov result Eq.~(\ref{EMFLHYscalar}), which is $\eta \ll 1 $, equivalent to saying that there should be many particles in the healing volume $\xi^D$. That the standard perturbation theory has a problem in the thermodynamic limit can be understood by noting that the integral in Eq.~(\ref{g3}) diverges at small $k$. The Bogoliubov theory cures this divergence effectively cutting off this integral in the infrared at momentum $1/\xi\sim \sqrt{Un}$. One can thus regard the LHY term as the three-body term with density dependent $g_3$.

We can name a few cases where the leading BMF correction is in the form of a three-body term in the thermodynamic limit. In particular, this happens for a special type of interaction for which $U(0)$ vanishes or is very small. The two-body interaction is then weak and the system is in the interesting for us regime where MF and LHY terms can be comparable to each other. In Secs.~\ref{Sec:polarons} and \ref{Sec:dipoles} we present two examples, originally discussed in Ref.~\cite{PricoupenkoPetrov2021}, for which Eq.~(\ref{g3}) becomes useful. 

\subsection{Three-body interaction of polarons}\label{Sec:polarons}

Let us go back to the Bose-Bose mixture discussed in Sec.~\ref{SecBoseBoseMixture} and consider its extreme mass- and population-imbalanced limit. Namely, we take three atoms of species $\2$ in a Bose-Einstein condensate (host gas) of species $\1$ \cite{Stoof2000,Viverit2000} assuming that $m_\2=1\gg m_\1$ so that we can integrate out the host-gas dynamics in the adiabatic Born-Oppenheimer approximation. In this manner the phonon exchange in the host gas leads to a static induced Yukawa attraction between the impurities and their direct interaction can be tuned in order to reach the condition $U(0)=0$. This is attained at the phase-separation threshold $g_{\1\2}=\sqrt{g_{\1\1}g_{\2\2}}$. Indeed, the Fourier transform of the effective interaction (exchange plus direct) between two $\2$ impurities is given by
\begin{equation}\label{Yukawa}
U({\bf k})=g_{\2\2}-\frac{g_{\1\2}^2/(g_{\1\1}\xi^2)}{k^2 + 1/\xi^2},
\end{equation}
where $\xi = 1/\sqrt{4m_\1 g_{\1\1} n_\1}$ \cite{Stoof2000,Viverit2000}.

Substituting Eq.~(\ref{Yukawa}) into Eq.~(\ref{g3}) leads to the three-body coupling constant
\begin{equation}\label{g3Yukawa}
g_3 = S_Du^3\xi^{4-D},
\end{equation}
where $u = g_{\2\2}=g_{\1\2}^2/g_{\1\1}$, $S_1=3/8$, $S_2=3/(4\pi)$ and $S_3=9/(16\pi)$. 

This result can be used, for instance, to study the mixture just above the threshold for the phase separation, where $g_{\2\2}$ becomes slightly below $g_{\1\2}^2/g_{\1\1}$. The polaron gas is then characterized by attractive two-body ($g_2=g_{\2\2}-g_{\1\2}^2/g_{\1\1}$) and repulsive three-body interactions with the energy density Eq.~(\ref{TwoThreeBodyEnergy}) realizing the setup suggested in Ref.~\cite{Bulgac} for observing quantum droplets. One can show that the liquefaction of polarons in this system and under considered conditions ($m_\2\gg m_\1$) is directly related to the phenomenon of partial miscibility~\cite{NaidonPetrov}. 

\subsection{Quasi-two-dimensional dipolar bosons}\label{Sec:dipoles}

\begin{figure}[hbtp]
\begin{center}
\includegraphics[clip,width=1\columnwidth]{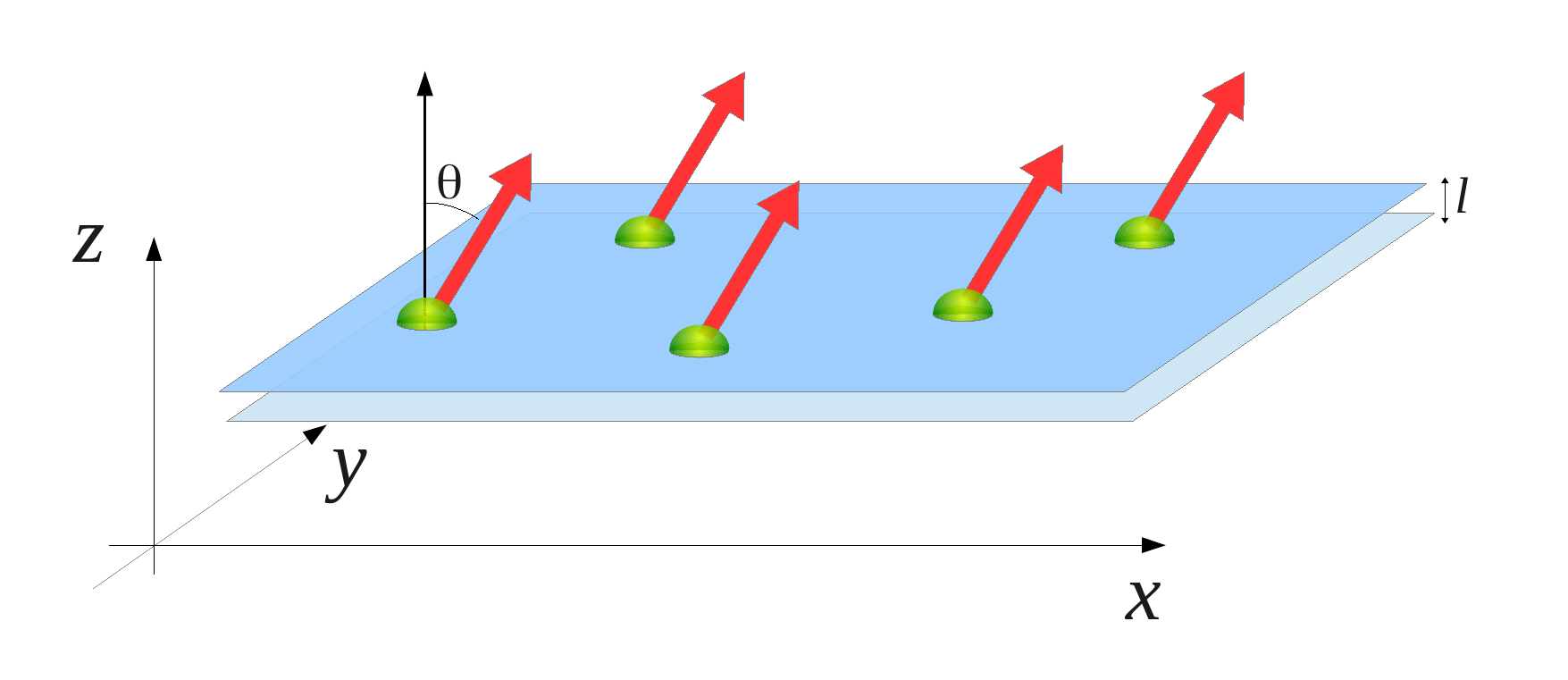}
\end{center}
\caption{
Schematic representation of a quasi-two-dimensional Bose gas of tilted dipoles.}
\label{FigTilted2Ddipoles}
\end{figure}

Another physically interesting case where Eq.~(\ref{g3}) can be used to estimate the leading BMF term is quasi-two-dimensional dipoles. Consider particles of unit mass interacting with each other by the (pseudo)potential~\cite{YiYou2000} 
\begin{equation}\label{dipolarV}
U({\bf r})=r_* \frac{r^2-3(x\sin\theta+z\cos\theta)^2}{r^5}+4\pi a \delta({\bf r})
\end{equation}
under the external confinement in the $z$-direction
\begin{equation}\label{ExtPot}
V(z)=\frac{z^2}{2l^4},
\end{equation}
where $l$ is the confinement oscillator length. The dipole moments are tilted by the angle $\theta$ with respect to the $z$-axis (see Fig.~\ref{FigTilted2Ddipoles}).

Let us pass from the three-dimensional system to the two-dimensional model by projecting the interaction potential onto the ground transversal eigenstate $\psi_0(z)=e^{-z^2/(2l^2)}/\sqrt{l\sqrt{\pi}}$. The Fourier transform of the resulting two-dimensional potential reads
\begin{equation}\label{Quasi2DPot}
U({\bf k})=2\sqrt{2\pi}\frac{a-(1/3-\cos^2\theta)r_*}{l}-2\pi r_* k e^{k^2l^2/2}{\rm erfc}(kl/\sqrt{2})[\cos^2\theta-(k_x/k)^2\sin^2\theta],
\end{equation}
where ${\rm erfc}(\sigma)$ is the complementary error function. We are interested in the vicinity of the point $U(0)=0$ which occurs when $a=(1/3-\cos^2\theta)r_*$. Near this point, substituting Eq.~(\ref{Quasi2DPot}) into Eq.~(\ref{g3}) and integrating over ${\bf k}$ gives
\begin{equation}\label{g3Quasi2D}
g_3=-127.4\frac{1+3\cos 2\theta}{4}\frac{31+12\cos 2\theta+21\cos 4\theta}{64}\frac{r_*^3}{l}.
\end{equation}
Equation~(\ref{g3Quasi2D}) is valid when $\{|a|,|r_*|\}\ll l$. Assuming $r_*>0$ we obtain that the effective three-body interaction is attractive for dipoles aligned along the $z$-direction (for $\theta=0$ we have $g_3=-127.4r_*^3/l$) and repulsive when they are parallel to the plane ($g_3=39.8r_*^3/l$ for $\theta=\pi/2$). 

The three-body term (\ref{g3Quasi2D}) represents the leading-order BMF effect in the dilute limit. From the form of the potential Eq.~(\ref{Quasi2DPot}) we see that the main contribution to the integral in Eq.~(\ref{g3}) comes from momenta $k\sim 1/l$. Therefore, in an effective low-energy theory for momenta $k\ll 1/l$ this terms can be included as a local three-body potential $g_3n^3/6$. Clearly, this term is important when $g_3 n$ becomes comparable to $U({\bf k})$ at a certain characteristic momentum for the system at hand (inverse system size, modulation momentum in the stripe phase, etc.) From Eq.~(\ref{Quasi2DPot}) we see that this happens for detunings $\delta a=a-(1/3-\cos^2\theta)r_*$ comparable to $r_*^3n$ and for momenta $kl\sim r_*^2n$, which can be relevant for studies of tilted quasi-two-dimensional dipoles in the dilute regime (see, for instance, Refs.~\cite{Fedorov2014,Baillie2015,Raghunandan2015}).

As we have just mentioned, the main contribution to the integral in Eq.~(\ref{g3}) comes from momenta $k\sim 1/l$. This means that this term accounts for BMF processes for three particles when they are at a distance comparable to the oscillator length. We now see that projecting to the ground transversal state is illegal and we have to take into account transversally excited states. It turns out, however, that the contribution of these excited states is not very significant and Eq.~(\ref{g3Quasi2D}) remains qualitatively correct~\cite{PricoupenkoPetrov2021}. 

We should note that Zin and co-workers~\cite{Zin2021} also found an effective three-body repulsion in the same model at $\theta=\pi/2$ assuming, however, a periodic box in $z$ direction instead of the harmonic confinement Eq.~(\ref{ExtPot}).

\subsection{Rabi-coupled Bose-Bose mixture}\label{SecRC}

Another system where the LHY correction takes the form of a three-body term is a Rabi-coupled Bose-Bose mixture. More precisely, the LHY term in such a mixture crosses over from  analytic to nonanalytic behavior depending on the ratio of the Rabi frequency to the MF energy $\Omega/gn$. This happens because the Bogoliubov branch responsible for the LHY correction is gapped and the gap, proportional to the Rabi frequency $\Omega$, introduces a characteristic momentum $\sqrt{m\Omega}$, which effectively cuts off the divergence in Eq.~(\ref{EMFLHYscalarExp}). As this cutoff momentum is now density independent (in contrast to the inverse healing length), the LHY term expands in integer powers of $n$. 

The Bogoliubov analysis of Rabi-coupled mixtures is a straightforward generalization of the procedure described in Sec.~\ref{SecBoseBoseMixture}. Let us mention the key differences. The Hamiltonian of the Rabi-coupled mixture is obtained adding to Eq.~(\ref{Ham}) the coupling term
\begin{equation}\label{RabiHam}
-\frac{1}{2}\sum_{\bf k}\Omega\hat{a}_{\1,{\bf k}}^\dagger\hat{a}_{\2,{\bf k}}+\Omega^*\hat{a}_{\2,{\bf k}}^\dagger\hat{a}_{\1,{\bf k}},
\end{equation}
where we choose $\Omega$ to be real and positive (we can always do this by redefining the phases of the single-particle modes). The term (\ref{RabiHam}) can be realized, for example, by coupling two hyperfine states of an atom with a radio-frequency field or by placing particles in a bilayer or bitube geometry with nonvanishing interlayer or intertube tunneling. 

Assuming $\hat{a}_{\sigma,0}=\sqrt{n_\sigma}e^{i\phi_\sigma}$ the MF grand potential reads [cf. Eq.~(\ref{H0})]
\begin{equation}\label{H0Rabi}
H_0=-\Omega\sqrt{n_\1 n_\2}\cos(\phi_\1-\phi_\2)-\sum_\sigma \mu_\sigma n_{\sigma}+\frac{1}{2}\sum_{\sigma\sigma'}U_{\sigma\sigma'}(0)n_{\sigma} n_{\sigma'}.
\end{equation}
We immediately see that the coupling term reduces the $U(1)\times U(1)$ symmetry of the uncoupled mixture to $U(1)$ as it locks the phase difference. We thus expect only one gapless Bogoliubov branch (Goldstone mode) and the other one should be gapped. We proceed by setting $\phi_\1=\phi_\2=0$. The Bogoliubov matrix (\ref{BogMatrix}) now reads
\begin{equation}\label{BogMatrixRabi}
\cal{L}=\begin{pmatrix}
\hat{A} & \hat{B}\\
-\hat{B} & -\hat{A}
\end{pmatrix},
\end{equation}
where  
\begin{equation}\label{RabiA}
\hat{A}=\begin{pmatrix}
\frac{k^2}{2m_\1}+U_{\1\1}({\bf k})n_\1 +\frac{\Omega}{2}\sqrt{\frac{n_\2}{n_\1}}& U_{\1\2}({\bf k})\sqrt{n_\1 n_\2}-\frac{\Omega}{2}  \\
U_{\1\2}({\bf k})\sqrt{n_\1 n_\2}-\frac{\Omega}{2} & \frac{k^2}{2m_\2}+U_{\2\2}({\bf k})n_\2+\frac{\Omega}{2}\sqrt{\frac{n_\1}{n_\2}}
\end{pmatrix}
\end{equation}
and
\begin{equation}\label{RabiB}
\hat{B}=\begin{pmatrix}
U_{\1\1}({\bf k})n_\1 & U_{\1\2}({\bf k})\sqrt{n_\1 n_\2} \\
U_{\1\2}({\bf k})\sqrt{n_\1 n_\2} & U_{\2\2}({\bf k})n_\2
\end{pmatrix}.
\end{equation}
The ground state energy in the Bogoliubov approximation Eq.~(\ref{EMFLHY}) in the Rabi-coupled case becomes
\begin{equation}\label{EMFLHYRabi}
\begin{aligned}
E_{\rm MF+LHY}&=-\Omega\sqrt{n_\1n_\2}+\frac{1}{2}\sum_{\sigma\sigma'}U_{\sigma\sigma'}(0)n_{\sigma} n_{\sigma'}\\
&+\frac{1}{2}\sideset{}{'}\sum_{\bf k}\left[\sum_{s=\pm}E_{s,{\bf k}}-\sum_{\sigma}\left(\frac{k^2}{2m_{\sigma}}+U_{\sigma\sigma}({\bf k})n_\sigma+\sqrt{\frac{n_{\bar{\sigma}}}{n_\sigma}}\frac{\Omega}{2}\right)\right],
\end{aligned}
\end{equation}
where $\bar{\1}=\2$ and $\bar{\2}=\1$. The passage from the bare interaction $U$ to the renormalized $g$ is performed in the same manner as in Sec.~\ref{Subsec:Renorm}. For the LHY energy density of a three-dimensional Rabi-coupled mixture with short-range interactions we explicitly obtain~\footnote{Speaking about Rabi-coupled mixtures we usually assume $m_\1=m_\2$, but this is not obligatory. For instance, in a spin-dependent optical lattice $\1$ and $\2$ atoms can have different effective masses. One can also think about immersing them in a spin-dependent environment where they have generally different polaronic masses.}
\begin{equation}\label{LHYRabi}
E_{\rm LHY}=\int {\frac{d^3k}{2(2\pi)^3}} \left[\sum_{s=\pm}E_{s,{\bf k}}-\sum_{\sigma}\left(\frac{k^2}{2m_{\sigma}}+g_{\sigma\sigma}n_\sigma+\sqrt{\frac{n_{\bar{\sigma}}}{n_\sigma}}\frac{\Omega}{2}\right)+\sum_{\sigma\sigma'}\frac{2m_{\sigma\sigma'}g^2_{\sigma\sigma'}n_\sigma n_{\sigma'}}{k^2}\right].
\end{equation}

To make further discussion less technical let us consider the symmetric setup in three dimensions with $m_\1=m_\2=1$, $g_{\1\1}=g_{\2\2}=g>0$, and $\mu_\1=\mu_\2=\mu$. We are interested in the vicinity of the collapse threshold where $g_{\1\2}\approx -g$. In this case $n_\1=n_\2=n/2$ and the Bogoliubov modes read~\cite{Meystre}
\begin{eqnarray}
E_{-,{\bf k}}&=&\sqrt{c^2k^2+k^4/4},\label{EminusRC}\\
E_{+,{\bf k}}&=&\sqrt{(\Omega+k^2/2)[\Omega+(g-g_{\1\2})n+k^2/2]}\label{EplusRC},
\end{eqnarray}
where $c^2=(g+g_{\1\2})n/2$. This structure of the Bogoliubov modes is explained as follows. In the symmetric case we deal with a condensate of atoms in the internal state $(\ket{\1}+\ket{\2})/\sqrt{2}$. On the MF level the two-body interaction corresponds to the coupling constant $(g+g_{\1\2})/2$. The behavior of the lower branch Eq.~(\ref{EminusRC}) is thus clear. The upper branch Eq.~(\ref{EplusRC}) corresponds to excitations to the internal state $(\ket{\1}-\ket{\2})/\sqrt{2}$, which, in vacuum, is higher in energy by $\Omega$. 

In the interesting for us region of small $g+g_{\1\2}$ the MF term $(g+g_{\1\2}) n^2/4$ competes with the LHY correction, which can be approximated by its value at $g=-g_{\1\2}$:
\begin{equation}\label{LHYRabiPoint}
E_{\rm LHY}=\int {\frac{d^3k}{2(2\pi)^3}} [\sqrt{(\Omega+k^2/2)(\Omega+2gn+k^2/2)}-k^2/2-gn-\Omega+g^2n^2/k^2].
\end{equation}
For $\Omega\ll gn$, neglecting $\Omega$ in Eq.~(\ref{LHYRabiPoint}), we arrive at the usual nonanalytic behavior of the LHY term as in the uncoupled case [see Eqs.~(\ref{EMFLHYShortRange}) and (\ref{EMFLHYRes})]. In the opposite limit we can expand the integrand in Eq.~(\ref{LHYRabiPoint}) in terms of small $gn/\Omega$ obtaining [cf. Eq.~(\ref{EMFLHYscalarExp})]
\begin{equation}\label{LHYRabiExp}
E_{\rm LHY}=\int {\frac{d^3k}{(2\pi)^3}} \left[\frac{\Omega g^2}{(2\Omega+k^2)k^2}n^2+\frac{g^3}{(2\Omega+k^2)^2}n^3+...\right]=\frac{g^2\sqrt{\Omega}}{2\sqrt{2}\pi}\frac{n^2}{2}+\frac{3g^3}{4\sqrt{2}\pi\sqrt{\Omega}}\frac{n^3}{6}+...
\end{equation} 
Equation~(\ref{LHYRabiExp}) is an illustration of how to extract the three-body effective interaction from the many-body Bogoliubov treatment. Although the symmetric case can also be solved by using genuine few-body techniques~\cite{Petrov2014}, their application for nonsymmetric setups (with finite radio-frequency detuning $\mu_\1 \neq \mu_\2$, nonequal $g_{\1\1}\neq g_{\2\2}$, etc.) would require a substantial effort. By contrast, with the Bogoliubov approach we just deal with slightly more complicated expressions for $E_{\pm,{\bf k}}$, which could still be expanded at small $gn/\Omega$ and integrated over momenta, eventually producing $g_3$. An additional advantage of the Bogoliubov theory is that it can also be used everywhere in the crossover from small to large $gn/\Omega$. We should mention that exact few-body methods become useful (and necessary) if the condition $|g|\sqrt{\Omega}\ll 1$ is not satisfied. One can see this already on the two-body level. Indeed, the MF two-body interaction $\propto gn^2$ and its leading order LHY renormalization $\propto g^2\sqrt{\Omega}n^2$ [see Eq.~(\ref{LHYRabiExp})], are the first two terms in the expansion of the true two-body interaction shift in powers of $g\sqrt{\Omega}$. The Bogoliubov Hamiltonian cannot predict the next-order term for the same reason as it cannot reproduce the last term in Eq.~(\ref{g2}). The two-body scattering problem is solved in general in Ref.~\cite{Lavoine2021}. 

\subsection{Three-body effective interaction on the MF level}

Note that in all previously considered situations the effective three-body interaction emerges in the third order in $U$. In fact, it can also show up in the second order, but so far we have carefully avoided these cases. One reason is that the second-order three-body terms are always attractive whereas we are mostly interested in the repulsive case. Another reason is that the second-order three-body terms can be treated by properly solving the MF part of the problem, i.e., the Gross-Pitaevskii equation (or equations). These terms show up when the two-body interaction in the Hamiltonian can change the state of one atom without changing the state of the other. This is possible, for instance, in quasi-low-dimensional systems, where momentum is not conserved in all directions, or in systems with internal degrees of freedom, like Rabi-coupled mixtures just discussed.

To support this statement consider a system of bosons with the Hamiltonian 
\begin{equation}\label{HamManyBodyGen}
\hat{H}=\sum_{{\bm q},{\bm \nu}}\epsilon_{{\bm \nu},{\bm q}}\hat{a}_{{\bm q},{\bm \nu}}^\dagger\hat{a}_{{\bm q},{\bm \nu}}+\frac{1}{2}\sum_{{\bm q}_1,{\bm q}_2,{\bm k},{\bm \nu},{\bm \mu},{\bm \eta},{\bm \zeta}}U_{{\bm \mu}{\bm \nu}}^{{\bm \zeta}{\bm \eta}}({\bf k})\hat{a}_{{\bm q}_2+{\bm k},{\bm \mu}}^\dagger\hat{a}_{{\bm q}_1-{\bm k},{\bm \zeta}}^\dagger\hat{a}_{{\bm q}_2,{\bm \nu}}\hat{a}_{{\bm q}_1,{\bm \eta}},
\end{equation}
where the single-particle part has already been diagonalized. Greek indices may correspond to transversally excited states in quasi-low-dimensional geometries or, for the Rabi-coupled two-component mixture, to the lower and upper dressed single-particle states. An important difference between the interaction terms in Eqs.~(\ref{HamManyBodyGen}) and (\ref{Ubar}) is that in Eq.~(\ref{HamManyBodyGen}) matrix elements of the type $U_{{\bm 0}{\bm \nu}}^{{\bm 0}{\bm 0}}({\bf 0})$ do not vanish in general. This leads to the following phenomenon. In the absence of interactions all bosons condense in the ground state with ${\bm \nu}=0$ and ${\bf k}=0$. If we now try to construct the expansion Eqs.~(\ref{Expansion}-\ref{H4}) around this single-particle mode, we obtain an additional term, linear in creation and annihilation operators of noncondensed particles, 
\begin{equation}\label{H1}
\hat{H}_1=\frac{n^{3/2}}{2}\sideset{}{'}\sum_{\bm \nu}[U_{{\bm \nu}{\bm 0}}^{{\bm 0}{\bm 0}}({\bf 0})+U_{{\bm 0}{\bm 0}}^{{\bm \nu}{\bm 0}}({\bf 0})]\hat{a}_{{\bm 0},{\bm \nu}}^\dagger+[U_{{\bm 0}{\bm \nu}}^{{\bm 0}{\bm 0}}({\bf 0})+U_{{\bm 0}{\bm 0}}^{{\bm 0}{\bm \nu}}({\bf 0})]\hat{a}_{{\bm 0},{\bm \nu}}.
\end{equation}
This linear term is a consequence of the fact that the single-particle ground state ${\bm \nu}=0$, ${\bf k}=0$ is not the ground state of the classical interacting Hamiltonian. For instance, the classical GP solution for a Bose-Einstein condensate in a parabolic trap passes from the Gaussian to Thomas-Fermi limits as the ratio of the interaction per particle to the trap level spacing changes from small to large values. This is to say that the classical ground state involves more and more single-particle modes as the interaction or the number of atoms grows. Therefore, the energy of a confined condensate, already on the MF level, can develop contributions proportional to $N^3$. The same is true for the Rabi-coupled mixture; the polarization of the condensate (the composition of the upper and lower dressed states in the condensate wavefunction) can change on the MF level as the interaction or the total density grows. 

Nevertheless, constructing the perturbation theory around the mode ${\bm \nu}=0$, ${\bf k}=0$ is legal and may be convenient, for instance, if we are interested in the region close to the two-body zero crossing. The procedure is an alternative to perturbatively solving the GP equation in this limit. The leading-order three-body term emerges if we treat Eq.~(\ref{H1}) as a perturbation on top of the single-particle part of Hamiltonian (\ref{HamManyBodyGen}). The energy shift is obtained in the second order (the first-order term obviously vanishes). It equals $g_3n^3/3!$, where
\begin{equation}\label{g32}
g_3=-\frac{3}{2}\sideset{}{'}\sum_{{\bm \nu }} \frac{[U_{{\bm 0}{\bm \nu}}^{{\bm 0}{\bm 0}}({\bf 0})+U_{{\bm 0}{\bm 0}}^{{\bm 0}{\bm \nu}}({\bf 0})][U_{{\bm \nu}{\bm 0}}^{{\bm 0}{\bm 0}}({\bf 0})+U_{{\bm 0}{\bm 0}}^{{\bm \nu}{\bm 0}}({\bf 0})]}{\epsilon_{{\bm \nu}, {\bf 0}}-\epsilon_{{\bm 0}, {\bf 0}}}.
\end{equation}
Three-body energy shifts of this type  have been discussed in the context of quasi-one-dimensional~\cite{Muryshev2002,Mazets2008} and lattice~\cite{Tiesinga,Bloch} bosons where the interatomic interaction is a delta potential characterized by scattering length. The three-body shift is then weak compared to the leading two-body shift $U_{{\bm 0}{\bm 0}}^{{\bm 0}{\bm 0}}({\bf 0})n^2/2$. However, in the context of these lectures more interesting would be the case characterized by vanishing two-body interaction with finite $g_3$, implying vanishing diagonal matrix element $U_{{\bm 0}{\bm 0}}^{{\bm 0}{\bm 0}}({\bf 0})$ and finite off-diagonal matrix elements entering Eq.~(\ref{g32}). In order to fulfill these requirements the system should have a sufficient number of tunable parameters. One example is the Rabi-coupled mixture in the general case, for $g_{\1\1}\neq g_{\2\2}$ and/or $\mu_{\1}\neq \mu_{\2}$. Depending on the parameters one can explore regimes of repulsive BMF~\cite{Lavoine2021} or attractive MF~\cite{Hammond2022} three-body interactions in this system. Another setup where the three-body interaction remains finite at the two-body zero crossing is the quasi-one-dimensional gas of bosonic dipoles~\cite{Edler2017,PricoupenkoPetrov2021}. This system features a BMF three-body effective repulsion when the dipoles are aligned parallel to the axis of the confining trap, and for finite tilt angle the three-body potential is dominated by the attractive MF contribution.

\section{Conclusions}

In these lectures we discussed physical origins, physical consequences, and calculation methods for the leading BMF correction (the LHY term) to the energy of a weakly-interacting bosonic mixture. This is one of the two experimentally explored systems (the other is a Bose gas with dipolar interactions) where the leading MF term can be tuned to small values comparable to the formally subleading LHY correction. In this case, inherently quantum BMF effects come into play. A ``smoking gun'' for this BMF physics is the quantum stabilization of a collapsing gas leading to the formation of dilute droplets. The phenomenon of liquefaction in the dilute regime goes far beyond the microscopic van der Waals theory.

To describe the phenomenon of quantum stabilization we used a simple single-particle model with two degrees of freedom, unstable in the classical case and stable in the quantum case (therefore, the term quantum stabilization). We argued that this model is an analog of the Bose-Bose mixture beyond its collapse threshold. The total density plays the role of a classically unstable degree of freedom stabilized by quantum fluctuations in the spin degree of freedom. A significant part of the notes was then devoted to the derivation of the LHY term for the mixture in the Bogoliubov framework. 

We argued that the LHY term characterizes the Bogoliubov vacuum through the integral over zero-point energies of the Bogoliubov modes. Its functional dependence on the density thus strongly depends on various system parameters (dimensionality, shape of the interaction potentials, external confinement). The usual Bogoliubov excitation spectrum of the type $\sqrt{k^4+gnk^2}$ gives rise to a nonanalytic density dependence of the LHY term ($\propto n^{5/2}$ in three dimensions). However, if the branch is gapped (for instance, in a Rabi-coupled mixture) or if it is soft at low momenta (when the interaction potential vanishes at zero momentum), the corresponding contribution to the LHY correction becomes analytic and can be expanded in integer powers of the density: a two-body contribution proportional to $n^2$, a three-body contribution $\propto n^3$, etc. In this case, comparing the Bogoliubov hierarchy (MF level, quadratic Bogoliubov level, leading beyond-Bogoliubov level, etc.) with the few-body expansion in powers of the interaction strength, we establish to what extent the many-body Bogoliubov approach can be used for calculating few-body observables. The point is that, surprisingly, the Bogoliubov method turns out to be technically easier for this purpose. 

Finally, we dare to mention a couple of directions in which the theory described in these notes can be continued. If the MF solution probes only the bottom of the classical well, the LHY term explores much more the configurational space of the system, presumably providing more possibilities for engineering exotic effective low-energy theories. Obviously, to explore this direction we have to be able to calculate the LHY term for an arbitrary Hamiltonian and starting from a generally inhomogeneous condensate field $\Psi_0$. The resulting energy functional containing the MF and LHY terms has to be minimized with respect to $\Psi_0$. Braaten and Nieto~\cite{Braaten1997} considered a single-component trapped BEC with zero-range interactions and developed a gradient expansion for the LHY term valid when the characteristic size of spatial variations of $\Psi_0$ is much larger than the typical wavelength of excitations dominating the LHY integral. We think that their work can potentially be extended to inhomogeneous mixtures or other systems where the LHY term is relatively enhanced and where one can also expect the gradient terms to play an interesting role.

Another natural extension of this line of research is to go beyond the Bogoliubov approximation. Indeed, the LHY correction becomes important when we fine-tune the MF term to small values. We can now start thinking of more exotic Hamiltonians where both MF and LHY terms are made small thus emphasizing the next-order correction. As we have mentioned, in the three-dimensional case the leading beyond-Bogoliubov correction is nonuniversal in the sense that it depends on the shape of the interaction potential or, in the zero-range approximation, on a three-body parameter. However, in low-dimensional cases the beyond-Bogoliubov term is universal and is well-defined for Bose-Bose mixtures with zero-range interactions. Mora and Castin calculated the leading beyond-Bogoliubov correction to the energy of a homogeneous two-dimensional single-component Bose gas~\cite{MoraCastin2009}.

\acknowledgments

This work is supported by the ANR Grant Droplets No. ANR-19-CE30-0003-02. The author warmly thanks G. Astrakharchik, T. Bourdel, M. Fattori, G. Modugno, P. Naidon, A. Pricoupenko, A. Recati, L. Santos, G. Shlyapnikov, L. Tarruell, and M. Zaccanti for their direct or undirect participation in research project, on which these lectures are based. The author is also grateful to the Italian Physical Society and to the organizers of the summer school R. Grimm, M. Inguscio, and S. Stringari for setting up such a wonderful event.

\section{Appendix: Diagonalization of the Bogoliubov Hamiltonian}

The quadratic Bogoliubov part of a bosonic Hamiltonian, written through normally ordered products of creation and annihilation operators, has the form
\begin{equation}\label{AppH2nonsymm}
\hat{H}_2=\hat{a}^\dagger \hat{A} \hat{a} + \hat{a}^\dagger \hat{B} \hat{a}^\dagger + \hat{a} \hat{B}^* \hat{a},
\end{equation}
where $\hat{a}$ is a vector of bosonic field operators $\hat{a}_\nu$ annihilating particles in single-particle modes $\nu=1,..., N$, corresponding to internal (spin) and/or external (orbital) degrees of freedom. The $N\times N$ matrices $\hat{A}$ and $\hat{B}$ are, respectively, Hermitian and symmetric. Equation~(\ref{AppH2nonsymm}) can further be reduced to   
\begin{equation}\label{AppAH2}
\hat{H}_2=-\frac{1}{2}{\rm Tr}[\hat{A}]+\frac{1}{2}(\hat{a}^\dagger \hat{a})
\begin{pmatrix}
\hat{A} & \hat{B}\\
\hat{B}^* & \hat{A}^*
\end{pmatrix}
\begin{pmatrix}
\hat{a} \\
\hat{a}^\dagger
\end{pmatrix},
\end{equation}
where we use the commutation relation for bosonic $\hat{a}_\nu$ in order to make the quadratic term look like the standard oscillator Hamiltonian, which features $\hat{a}^\dagger \hat{a}$ and $\hat{a}\hat{a}^\dagger$ in a symmetric manner. The first term on the right-hand side of Eq.~(\ref{AppAH2}) thus compensates for the initial ``wrong'' ordering. The symmetrized matrix can formally be diagonalized by using eigenstates of the Bogoliubov matrix
\begin{equation}\label{L}
\cal{L}=\begin{pmatrix}
\hat{A} & \hat{B}\\
-\hat{B}^* & -\hat{A}^*
\end{pmatrix},
\end{equation}  
which itself emerges from the analysis of small-amplitude excitations of the classical analog of $\hat{H}_2$. 


$\cal{L}$ has the following properties.

Property 1: If
\begin{equation}\label{ABmBmAuv}
\begin{pmatrix}
\hat{A} & \hat{B}\\
-\hat{B}^* & -\hat{A}^*
\end{pmatrix}
\begin{pmatrix}
\vec{u}\\
\vec{v}
\end{pmatrix}=\epsilon
\begin{pmatrix}
\vec{u}\\
\vec{v}
\end{pmatrix},
\end{equation}
then
\begin{equation}\label{ABmBmAvu}
\begin{pmatrix}
\hat{A} & \hat{B}\\
-\hat{B}^* & -\hat{A}^*
\end{pmatrix}
\begin{pmatrix}
\vec{v}^*\\
\vec{u}^*
\end{pmatrix}=-\epsilon^*
\begin{pmatrix}
\vec{v}^*\\
\vec{u}^*
\end{pmatrix}.
\end{equation}
One consequence of this property is that if $\epsilon$ is real, there is another eigenstate with energy $-\epsilon$, Eq.~(\ref{ABmBmAvu}) giving the corresponding eigenvector. The real spectrum of (\ref{L}) is thus symmetric with respect to zero.


Property 2: If the right eigenvector of (\ref{L}) is defined by Eq.~(\ref{ABmBmAuv}), then $\begin{pmatrix}
\vec{u}^\dagger & -\vec{v}^\dagger \end{pmatrix}$ is the left eigenvector of (\ref{L})  corresponding to $\epsilon^*$, i.e.,  
\begin{equation}\label{ABmBmAuvLeft}
\begin{pmatrix}
\vec{u}^\dagger & -\vec{v}^\dagger
\end{pmatrix}
\begin{pmatrix}
\hat{A} & \hat{B}\\
-\hat{B}^* & -\hat{A}^*
\end{pmatrix}
=\epsilon^*
\begin{pmatrix}
\vec{u}^\dagger & -\vec{v}^\dagger
\end{pmatrix}.
\end{equation}
Again, if $\epsilon$ is real, Eq.~(\ref{ABmBmAuvLeft}) gives the left eigenvector corresponding to $\epsilon$. Left and right eigenvectors corresponding to different $\epsilon$ are orthogonal. This can be shown by multiplying $\cal{L}$ by its $i$-th left eigenvector from the left and by its $j$-th right eigenvector from the right (if the vectors were not orthogonal, we would have different results depending on what multiplication was performed first). 

Let us assume that the classical ground state is dynamically stable, i.e., the spectrum of (\ref{L}) is real. We normalize the states by requiring
\begin{equation}\label{NormBogoliubov}
\vec{u}^\dagger_i \vec{u}_j-\vec{v}^\dagger_i \vec{v}_j=\pm\delta_{i,j}.
\end{equation}
The sign distinguishes between the particle-like (+) and hole-like (-) state in the pair. Let us sort the eigenstates such that the first $N$ of them are particle-like states and the last $N$ are corresponding hole-like states. This is to say that, according to Property 1, the $N+j$-th eigenstate is obtained from the $j$-th one by exchanging $\vec{u}_j\leftrightarrow \vec{v}_j^*$. 
The matrix composed of the eigenvectors of $\cal{L}$ sorted in this manner 
\begin{equation}\label{Sformal}
\hat{S}=\begin{pmatrix}
\vec{u}_1&...&\vec{u}_N&\vec{v}_1^*&...&\vec{v}_N^*\\
\vec{v}_1&...&\vec{v}_N&\vec{u}_1^*&...&\vec{u}_N^*
\end{pmatrix}
\end{equation}
diagonalizes the matrix 
\begin{equation}\label{LbarABBA}
\bar{\cal{L}}=\begin{pmatrix}
\hat{A} & \hat{B}\\
\hat{B}^* & \hat{A}^*
\end{pmatrix}
\end{equation} 
as follows.
\begin{equation}\label{ABBAdiagonalized}
\hat{S}^\dagger
\begin{pmatrix}
\hat{A} & \hat{B}\\
\hat{B}^* & \hat{A}^*
\end{pmatrix}
\hat{S}=
\begin{pmatrix}
\epsilon_1 &&&&&0\\
&\ddots &&&&\\
&&\epsilon_N&&&\\
&&&\epsilon_1&&\\
&&&&\ddots&\\
0&&&&&\epsilon_N
\end{pmatrix}
\end{equation}
The Bogoliubov transformation to new modes is defined by
\begin{equation}\label{BogTransVec}
\begin{pmatrix}
\hat{a} \\
\hat{a}^\dagger
\end{pmatrix}
=\hat{S}
\begin{pmatrix}
\hat{b} \\
\hat{b}^\dagger
\end{pmatrix}.
\end{equation}
The inverse transformation is obtained by using the inverse matrix 
\begin{equation}\label{Sformalm1}
\hat{S}^{-1}=\begin{pmatrix}
\vec{u}_1^\dagger&-\vec{v}_1^\dagger\\
\vdotswithin{u}&\vdotswithin{v}\\
\vec{u}_N^\dagger&-\vec{v}_N^\dagger\\
-\vec{v}_1^T&\vec{u}_1^T\\
\vdotswithin{v}&\vdotswithin{u}\\
-\vec{v}_N^T&\vec{u}_N^T
\end{pmatrix},
\end{equation}which satisfies $\hat{S}^{-1}\hat{S}=1$. Using the orthogonality properties of the eigenstates one can show that the new operators $\hat{b}_\nu$ satisfy bosonic commutation relations.  

The Hamiltonian (\ref{AppH2nonsymm}) reduces to 
\begin{equation}\label{AppH2nonsymmDiag}
\hat{H}_2=E_{\rm LHY}+\sum_{\nu=1}^{N}\epsilon_\nu\hat{b}^\dagger_\nu \hat{b}_\nu,
\end{equation}
where the ground state energy (LHY correction) equals
\begin{equation}\label{AppEBMF}
E_{\rm LHY}=\sum_{\nu=1}^{N}\epsilon_\nu/2-\sum_{i=1}^NA_{ii}/2.
\end{equation}

The energies of particle-like eigenstates $\epsilon_\nu$ are not necessarily all positive. A simple example is a Bose gas in a moving reference frame with velocity ${\bf v}$ larger than the speed of sound. Negative energies in this case point to a thermodynamic instability. Interestingly, the BMF correction in this case does not depend on ${\bf v}$ as the shift of the Bogoliubov energies is compensated by the shift of $A_{ii}$. In all examples discussed in these lectures particle-like excitations have positive energies.

In general, the spectrum of (\ref{L}) contains complex frequencies. They can come either as positive and negative purely imaginary pairs [see below Eq.~(\ref{Haaaa}) corresponding to $\hat{A}=0$, $\hat{B}=-1$] or quartets of the type $\pm\epsilon$, $\pm\epsilon^*$ [check $\hat{A}= \begin{pmatrix}
1 & 0\\
0 & 2
\end{pmatrix}$ and $\hat{B}= \begin{pmatrix}
0 & 2i\\
2i & 0
\end{pmatrix}$]. The subset of excitations with real frequences can still be diagonalized as shown above (i.e., they reduce to a sum of terms of the type ${\hat b}^\dagger_\nu\hat{b}_\nu$). One can also show that they are decoupled from the excitations with complex frequencies, which themselves can be diagonalized in a block-diagonal manner, where each block corresponds either to a purely imaginary complex-conjugate pair of frequences or to a quartet of the type $\pm\epsilon$, $\pm\epsilon^*$. However, these blocks cannot be diagonalized. An illustrative example is the expulsive harmonic oscillator $\hat{p}^2/2-\hat{x}^2/2$. By using $\hat{a}=\hat{x}/\sqrt{2}+i\hat{p}/\sqrt{2}$ and $\hat{a}^\dagger=\hat{x}/\sqrt{2}-i\hat{p}/\sqrt{2}$ we can write its Hamiltoninan in the form
\begin{equation}\label{Haaaa}
\hat{H}=-\frac{\hat{a}\hat{a}+\hat{a}^\dagger\hat{a}^\dagger}{2}.
\end{equation} 
The corresponding Bogoliubov frequencies are $\pm i$. It can be shown by direct substitution that there is no Bogoliubov transformation (preserving bosonic commutation)
\begin{eqnarray}\label{Bogaaaa}
\hat{a}&=&u\hat{b}+v^*\hat{b}^\dagger,\nonumber\\
\hat{a}^\dagger&=&u^*\hat{b}^\dagger+v\hat{b},
\end{eqnarray}
that can reduce Eq.~(\ref{Haaaa}) to the form $\propto \hat{b}^\dagger\hat{b}+{\rm const}$.

\end{document}